\documentclass{article}
\usepackage{amsmath,amssymb}
\usepackage{mathtools}
\usepackage{amsfonts}
\usepackage{color}
\usepackage{moreverb}
\usepackage{graphicx}
\usepackage{indentfirst}
\usepackage{booktabs}
\usepackage{rotating}
\usepackage{verbatim}
\usepackage{float} 
\usepackage{breqn}
\usepackage{caption}
\usepackage{subcaption}

\newcommand{\vect}[1]{\boldsymbol{#1}}

\newcommand{\mat}[1]{\mathbf{#1}} 
\newcommand{\A}{\mathbf A}
\newcommand{\bb}{\boldsymbol b}

\newcommand{\e}{\vect e}
\newcommand{\n}{\boldsymbol n}

\newcommand{\bO}{\mat 0}
\newcommand{\bt}{{\boldsymbol t}}
\newcommand{\btt}{\boldsymbol{\tilde t}}

\newcommand{\bxi}{{\boldsymbol\xi}}

\newcommand{\btheta}{\vect{\theta}}
                               
\newcommand{\by}{\boldsymbol y}

\newcommand{\g}{{\boldsymbol g}}
\newcommand{\gt}{\tilde{\boldsymbol g}}

\newcommand{\J}{\mat J}
\newcommand{\K}{\mat K}

\newcommand{\br}{\boldsymbol r}

\newcommand{\veps}{\varepsilon}
\newcommand{\T}{\boldsymbol T}
\newcommand{\X}{\boldsymbol X}
\newcommand{\x}{\boldsymbol x}
\newcommand{\y}{\boldsymbol y}

\DeclareMathOperator*{\argmin}{argmin}

\begin{document}

\title{Hybrid cell-centred/vertex model for multicellular systems with equilibrium-preserving remodelling}
\author{P. Mosaffa, A. Rodr\'iguez-Ferran, J.J. Mu\~noz (\tt{j.munoz@upc.edu})\\
\centerline{\emph{Unviersitat Politècnica de Catalunya, Barcelona, Spain}}
}



\maketitle
\begin{abstract}
We present a hybrid vertex/cell-centred model for mechanically simulating planar cellular monolayers undergoing cell reorganisation.  Cell centres are represented by a triangular nodal network, while the cell boundaries are formed by an associated vertex network. The two networks are coupled through a kinematic constraint which we allow to relax progressively. Special attention is paid to the change of cell-cell connectivity due to cell reorganisation or remodelling events. We handle these  situations by using a variable resting length and applying an Equilibrium-Preserving Mapping (EPM) on the new connectivity, which computes  a new set of resting lengths that preserve nodal and vertex equilibrium. We illustrate the properties of the model by simulating monolayers subjected to imposed extension and during a wound healing process. The evolution of forces and the EPM are analysed during the remodelling events. As a by-product, the proposed technique enables to recover fully vertex or fully cell-centred models in a seamlessly manner by modifying a numerical parameter of the model.
\end{abstract}

\textbf{keywords}: cell-centred, vertex model, remodelling, tessellation, biomechanics, tissues.

\section{Introduction}

Mechanical analysis of embryonic tissues has gained attention in recent years. Biologists and experimentalists have been able to accurately track the kinematic information of tissues and organs, but the mechanical forces that drive these shape changes have resulted far more elusive, despite evidence that genetic expression and mechanics are tightly coupled in cell migration \cite{sunyer15}, wound healing \cite{brugues14} or embryo development \cite{fernandez15}. 

The quantification of the  mechanical forces in morphogenesis has given rise to numerous and diverse numerical approaches \cite{wyczalkowski12}, which can be classified into two main types: continuum and cell-based models. The former allow to incorporate well-known constitutive behaviour of solids or fluids \cite{bowden16} and can be discretised with robust techniques such as finite elements \cite{conte08,menzel12}. The latter instead have the advantage of explicitly representing junctional mechanics and capturing the discrete and cellular nature of tissues \cite{davidson10,hardin04,perrone16}. Due to recent evidence on the role of contractile forces at cell-cell junctions in embryonic development \cite{munjal15} and wound healing \cite{vedula15}, we will here present a methodology based on the latter approach.

Cell-based models can be described through cell-centred or off-latice models \cite{drasdo05,mirams13,pathmanathan09,vermolen12}, or vertex models (see for instance \cite{honda83,krajnc13,okuda15,weliky90} and the review articles \cite{alt17,fletcher13}). The first approach focuses on establishing forces between cell-centres and can easily include variations on the number of cells (cell proliferation or apoptosis). The second approach is instead driven by the mechanical forces at the cell-cell junctions \cite{schilling11}, which seem to determine the emergent properties of tissues and monolayers \cite{harris12}.  

The model proposed here aims to gather the advantages of the two approaches: define cell-cell interactions between centres and at the cell-cell junctions, but include the cell as an essential unit in order to ease the transitions in the cell-cell contacts. We resort to Delaunay triangulation of the cell-centres, and a barycentric interpolation of the vertices on the cell-boundaries. Both nodes and vertices are kinematically coupled by this interpolation, which has effects on the resulting equilibrium equations. 

The use of Voronoi tessellations has been well studied for domain decomposition \cite{fu17} or for discretising partial differential equations in elasticity, diffusion, fluid dynamics or electrostatics. Some examples are the Natural Element  Method \cite{cueto02,sibson80,sukumar03}, the Voronoi Cell Finite Element method \cite{moorthy96}, the Voronoi Interface Element \cite{guittet15} or the particle-in-cell methodology \cite{gatsonis09}. In these methods, the tessellation is used for either constructing the interpolation functions, or describing the heterogeneities or interfaces. 

We resort here to the related barycentric tessellation, where the vertices of the network are built from the barycentres of each triangle instead of the bisectors, as it is the case in the Voronoi diagram. We choose this alternative tessellation to guarantee that the vertices are inside each triangle, even when the Delaunay triangulation is deformed, and thus may potentially violate the Delaunay condition. The use of automatic tessellation is also motivated in our case by the need to handle cell-cell connectivity changes in a robust and accurate manner, and thus avoid the design of specific algorithms during remodelling events, as it is customary in vertex models in two \cite{fletcher13,honda83,schilling11} and three dimensions \cite{honda08,okuda15}. 

The proposed model extends a previous cell-centred model \cite{mosaffa15} with a hybrid approach that incorporates mechanics at the cell boundaries in order to model morphogenetic events driven by contractile forces \cite{salbreux12b}, like for instance germ band extension \cite{munjal15} or wound healing \cite{antunes13}. Other recent hybrid techniques that couple cell-centred and continuum approaches may be found in \cite{gonzalez17}, but with no specific mechanics at the cell junctions.

We point out that our aim is to be able to model multicellular systems, with hundreds of cells. We therefore focus our approach at the cell rather than at the subcellular scale. Other methods for modelling cell mechanics such as the Subcellular Element Model \cite{sandersius08,sandersius11} or the Immersed Boundary Method \cite{rejniak07} are more suitable at smaller scales and therefore can simulate cell-cell interaction more accurately. 
  
We will first define the model kinematics in Section \ref{s:kine} and 
the equations that describe the mechanical equilibrium of the multicellular system in Section \ref{s:mecha}. The particular viscoelastic rheological model is presented in Section \ref{s:rheo}; it allows to handle inter-cellular remodelling by using the equilibrium-preserving mapping described in Section \ref{s:map}. Representative results are presented in Section \ref{s:results} and some conclusions are highlighted in Section \ref{s:conclusions}.

\section{Tissue discretisation}\label{s:kine}

\subsection{Nodal and vertex networks}

\begin{figure}[!htb]
\centering
\includegraphics[scale=0.3]{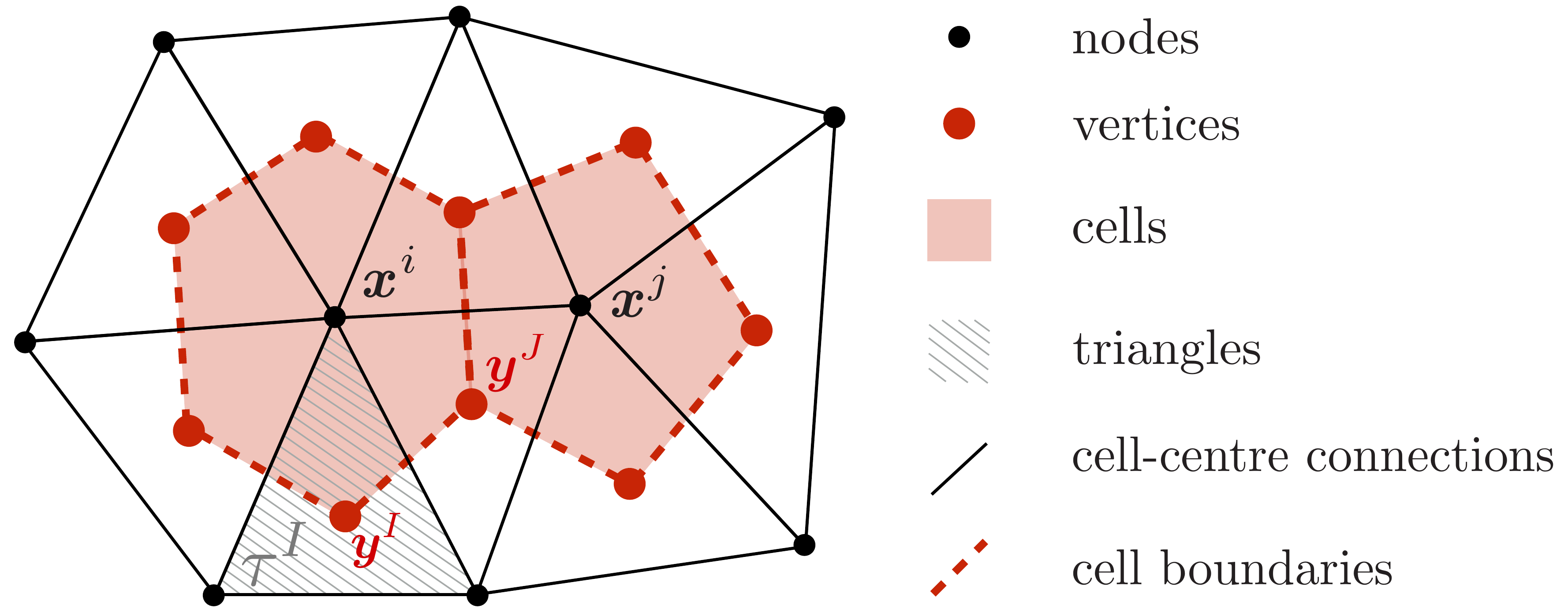}
\caption{Discretisation of tissue into cell-centres (nodes, $\x^i$) and cell boundaries (vertices, $\y^I$). Nodal network and vertex network are outlined with continuous and dashed lines, respectively. } 
\label{f:geometry}
\end{figure}

In the proposed model the tissue kinematics is defined by the cell-centres or \emph{nodes} and the cell boundaries, which are formed by a set of \emph{vertices}. We will denote by $\x^i$ the nodal positions (lower case superscript), and by  $\y^I$ the vertex positions (upper case superscript). In \ref{s:notation} we give a complete list of the notation employed in the article. Figure \ref{f:geometry} shows an example of the nodal and vertex networks that define the domain of a tissue. The bar elements that define each one of the networks will be in turn employed to write the mechanical equilibrium equation. In the next subsections we detail the definitions of the nodal and vertex positions and their relation. 

\subsection{Nodal geometry}\label{s:ngeo}

We will assume that a tissue forms a flat surface and has a constant number of nodes $N_{nodes}$. These are kinematically described by their cell-centres positions $\X= \left\{\x^1,...,\x^{N_{nodes}}\right\}$ and connectivity $\T$, which define a triangulation of the domain into $N_{tri}$ triangles $\mathcal T^I, I=1,\ldots, N_{tri}$ and $N_{D}$ edges. We will denote by $\X_n$ and $\T_n$ the set of nodal coordinates and connectivity at time $t_n$. Figure \ref{f:geometry} illustrates the connectivity of the nodal network.

\begin{figure}[!htb]
\centering
\includegraphics[width=1.0\textwidth]{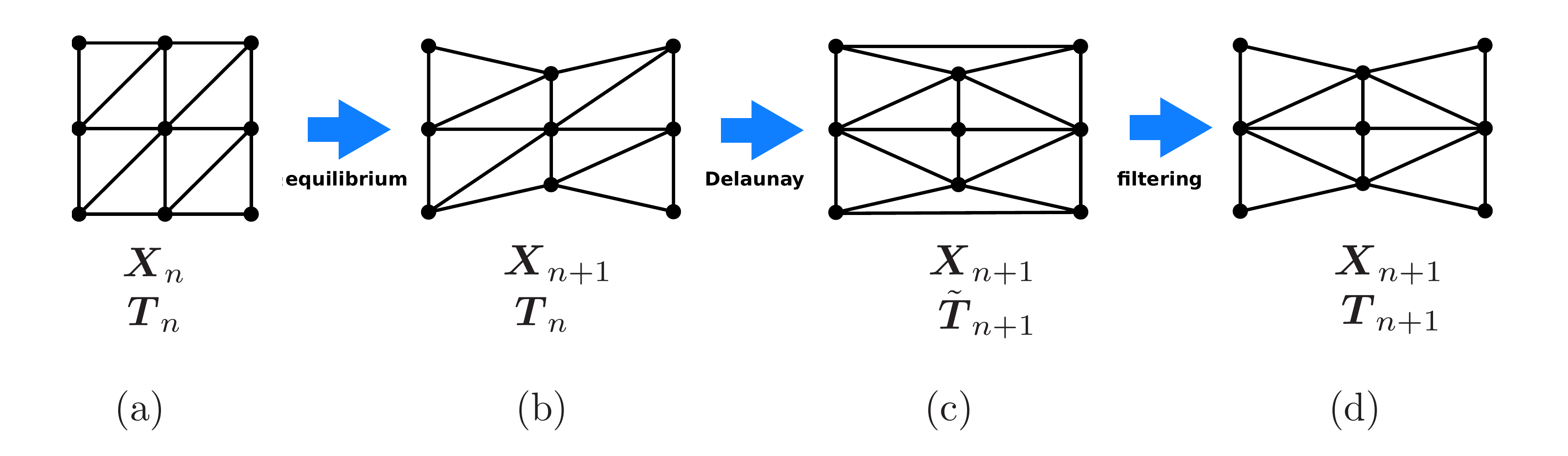}
\caption{Schematic of computational process for retrieving nodal positions and connectivity $\left\{\X_{n+1}, \T_{n+1}\right\}$ from the same quantities at time $t_n$. (a)$\rightarrow$(b): computation of new positions $\X_{n+1}$ from mechanical equilibrium. (b)$\rightarrow$(c): computation of new connectivity $\tilde\T_{n+1}$ from Delaunay triangulation. (c)$\rightarrow$(d): trimming of Delaunay connectivity $\tilde\T_{n+1}$, resulting in a not necessarily convex boundary of the cell-centred network $\T_{n+1}$.}
\label{f:steps}
\end{figure}

The position of the nodes is resolved using mechanical equilibrium, which will be explained in Section \ref{s:mecha}. The connectivities are found resorting to a trimmed Delaunay triangulation in order to obtain a not necessarily convex boundary. Triangles with an aspect ratio larger than a given tolerance are removed, and each pair of connected nodes $\x^i$ and $\x^j$ are connected with a bar element, with a rheology that will be detailed later. Figure \ref{f:steps} illustrates this trimming process and the steps for obtaining configuration $\{\X_{n+1}, \T_{n+1}\}$ from $\{\X_n, \T_ n\}$.

\subsection{Vertex geometry}\label{s:vertex}
 
The boundaries of the cells are defined by a set of connected vertices  $\left\{\y^1,\ldots,\right.$ $\left.\y^{N_{tri}}\right\}$, which define a tessellation of the tissue domain into $\bar N_{nodes}$ cell domains $\Omega^i$, $i=1,\ldots, \bar N_{nodes}$. Note that $\bar N_{nodes}< N_{nodes}$ because $\bar N_{nodes}$ does not include the external nodes. Each triangle $\mathcal T^I$ is associated to vertex $\y^I$, and each interior node $i$ is surrounded by a number of vertices which is not necessarily constant between time-steps and may vary from cell to cell (see Figure \ref{f:geometry}). 

The position of vertex $\y^I$ is given by a local parametric coordinate $\bxi^I$ in triangle $\mathcal T^I$. The kinematic relation between the nodal positions $\x^i$ and the vertices is given by the interpolation
\begin{align}\label{e:interp}
\y^I=\sum_{i\in \mathcal T^I} p^i(\bxi^I)\x^i.
\end{align}

The previous summation extends to the three nodes of triangle $\mathcal T^I$ where vertex $I$ is located. Function $p^i(\bxi^I)$ is the standard finite element interpolation function of node $i$ in triangle $\mathcal T^I$ evaluated at coordinate $\bxi^I$. We will initially consider that all parameters $\bxi^I$ have a constant value $\bxi^I=\frac{1}{3}\{1\ 1\}$, which  corresponds to a barycentric tessellation of the domain. We will eventually allow varying values of $\bxi^I$ in Section \ref{s:xirelax}, where $\xi$-relaxation is introduced. 

Every two vertices $\y^I$ and $\y^J$ are connected with a bar element if their corresponding triangles $\mathcal T^I$ and $\mathcal T^J$ have a common edge. The positions and the connectivity of nodes and vertices in the tessellated network is uniquely defined by $\X$, $\T$, and all the local coordinates $\bxi=\{\bxi^1, \ldots, \bxi^{N_{tri}}\}$ which define the vertex locations $\y^I, I=1,\ldots, N_{tri}$. The rheology of  the $N_V$ bar elements that join the vertices along the boundary of cells will be also described in Section \ref{s:rheo}. 

We remark that the Voronoi tessellation of the tissue may be obtained by computing  specific  values of the parameter $\bxi^I$ for each vertex. However, we will not consider this tessellation in this article because our initial Delaunay triangulation deforms due to mechanical equilibrium, with a potential loss of its Delaunay character. In this case, Voronoi tessellation may become undefined, or lead to crossing bars or overlapping domains. 

\section{Mechanical equilibrium}\label{s:mecha}

Mechanical equilibrium of the bar elements that form the nodal and vertex networks is computed by minimising the total elastic energy of the two networks. This energy is decomposed as the sum of a nodal contribution $W_D(\x)$, and a contribution of the vertex network, $W_V(\by(\x))$. The minimisation of the total elastic energy $W_D(\x) + W_V(\by(\x))$ with respect to the nodal positions in $\X$, which are considered the principal kinematic variables, yields the equations
\[
\frac{\partial W_D(\x)}{\partial \x^i} + \frac{\partial W_V(\by(\x))}{\partial \x^i}=\bO,\  i=1,\ldots, N_{nodes}
\]

We will consider each one of the two terms on the left separately in the next subsections.

\begin{figure}[!htb]
\centerline{
\subcaptionbox{\label{f:nodesvertexA}}{\includegraphics[width=.35\textwidth]{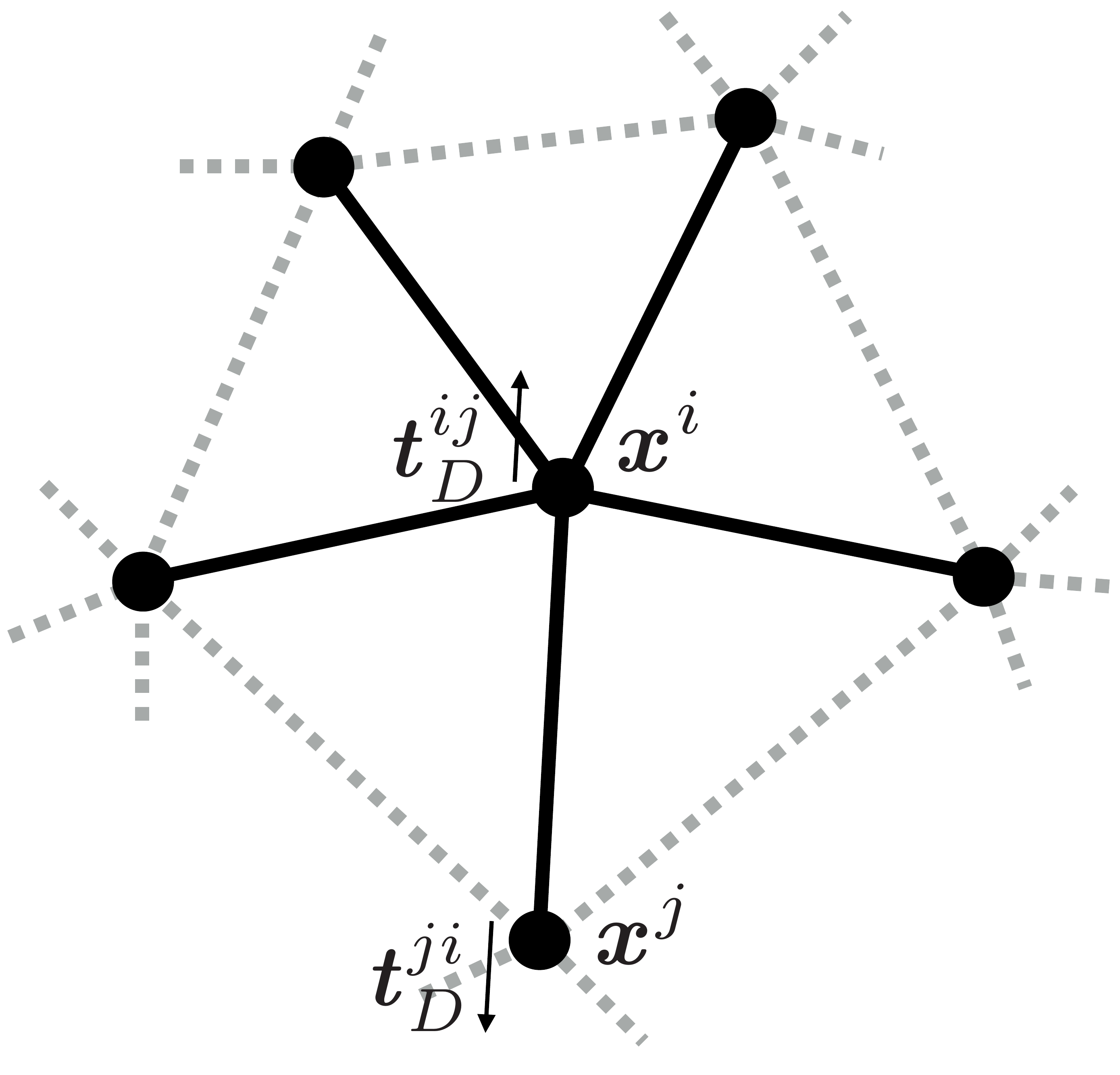}}
  \hspace{5ex}
\subcaptionbox{\label{f:nodesvertexB}}{\includegraphics[width=.35\textwidth]{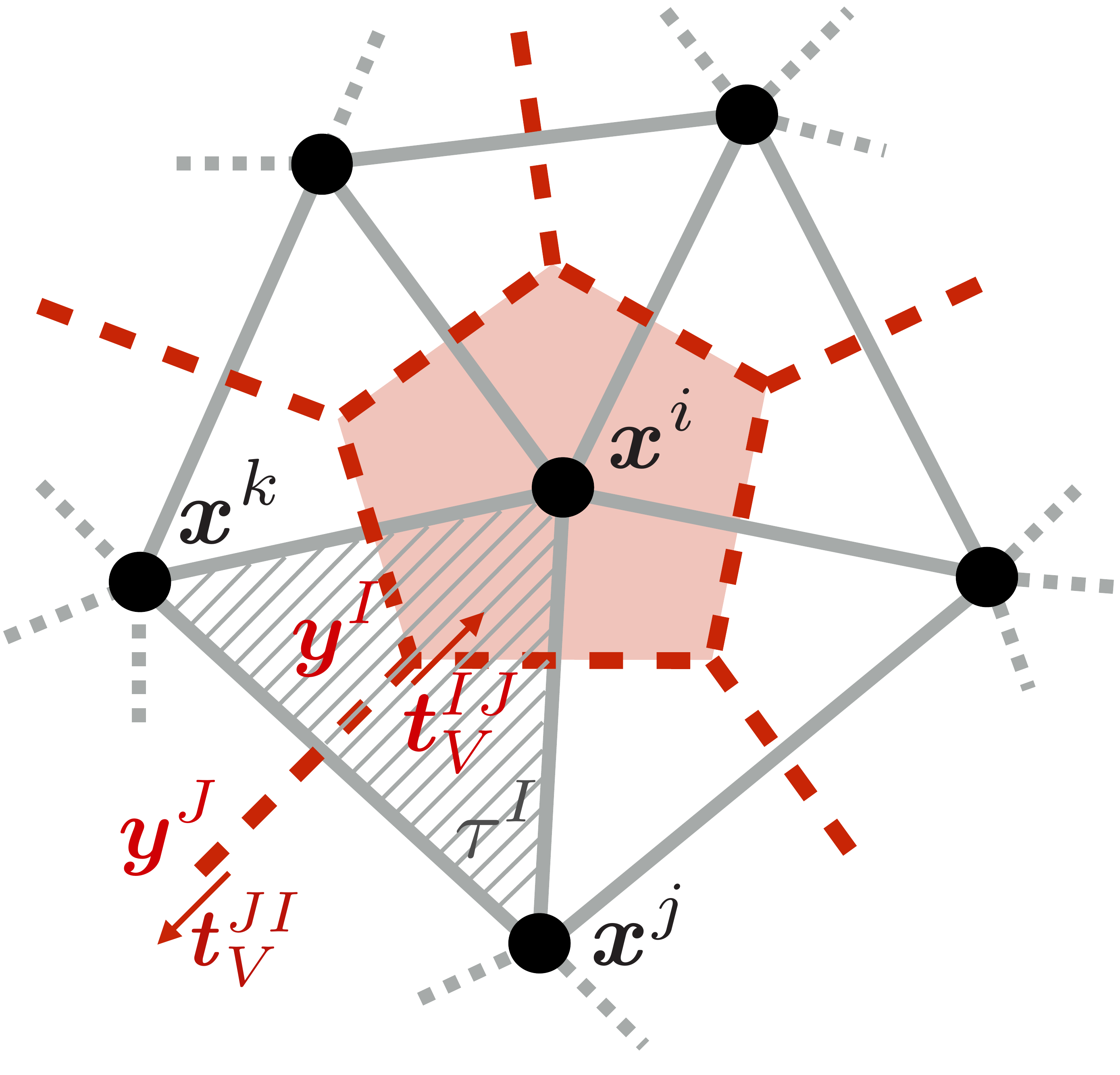}}
} 
\caption{(a): Schematic view of node $i$ connectivity (continuous lines), within  the rest of the network (dashed lines) and traction vector $\bt^{ij}_D$. (b): Cell boundary (highlighted polygon) corresponding to node $i$. Barycentric tessellation of triangle $ijk$ results to triple-junction $\y^I$. Vector $\bt_V^{IJ}$ represents the traction between vertices $\y^I$ and $\y^J$ along the shared boundary of cells $\x^j$ and $\x^k$.}
\label{f:nodesvertex}
\end{figure}

\subsection{Cell-centred mechanical equilibrium}

The cell-cell connectivity defined by $\T$ includes information on the set of $N_D$ pairs $ij$ between the $N_{nodes}$ nodes. Each pair of connected nodes are joined with a bar element that represents the forces between the two cells. This force is derived here from an elastic strain function,
\begin{align}\label{e:W_D}
\begin{split}
W^{ij}_D(\x)&=\frac{1}{2}k_D(\veps^{ij})^2,\\
W_D(\x)&=\sum_{ij=1}^{N_D}W^{ij}_D(\x),
\end{split}
\end{align}
where $k_D$ is the material inter-cellular stiffness, $\varepsilon^{ij}=\frac{l^{ij}-L^{ij}}{L^{ij}}$ is the scalar elastic strain, and $l^{ij}=\left\|\x^i-\x^j\right\|$ and $L^{ij}$ are the \textit{current} and \textit{reference lengths}. In Section \ref{s:rheo} we will introduce a rheological law where the reference length $L^{ij}$ (stress-free length of the element) is allowed to vary along time, and thus we may have that $L^{ij}\neq L_0^{ij}:=\left\|\x_0^i-\x_0^j\right\|$. $W_D$ is the total strain function of the network of nodes. In the absence of any other strain function, the minimisation of $W_D$ leads to the equations
\begin{align}\label{e:g_D}
\g_D^i:=\sum_{j\in{S^i}}\bt^{ij}_D=\bO,\ i=1,\ldots,N_{nodes},
\end{align}
where $S^i$ denotes the set of nodes connected to node $i$ and $\bt^{ij}_D$ is the nodal traction at node $i$ due to bar $ij$, which is derived from the elastic strain function $W^{ij}_D$ as (no summation on $i$)
\begin{align}\label{e:td}
\bt^{ij}_D=\frac{\partial W^{ij}_D}{\partial \x^i}=-\bt^{ji}_D=-\frac{\partial W^{ij}_D}{\partial \x^j}.
\end{align}

Figure \ref{f:nodesvertexA} shows the traction vectors between two nodes $\x^i$ and $\x^j$. Since the system of equations \eqref{e:g_D} is non-linear with respect to the nodal positions
$\x^i$, we resort to a full Newton-Raphson method, which requires linearisation of the set of equations. The expression of the resulting Jacobian is given in \ref{s:linearisation}.

\subsection{Adding vertex mechanical equilibrium}\label{s:mvertex}

The force between any two vertices is also derived here from an elastic strain function,
\begin{align}\label{e:W_V}
\begin{split}
W^{IJ}_V(\by)=\frac{1}{2}k_V(\veps^{IJ})^2\\
W_V(\by)=\sum_{IJ=1}^{N_V}W_V^{IJ}(\by)
\end{split}
\end{align}
with $k_V$ the cell boundary stretching stiffness. The total mechanical strain energy of the system is the sum of the contributions of the nodal and vertex networks,
\begin{align*}
W_D(\x)+W_V(\by(\x)).
\end{align*}

The new nodal positions are found by solving the minimisation problem
\begin{align}\label{e:minix}
\x^*=\argmin_\x \Big(W_D(\x) + W_V(\by(\x))\Big).
\end{align}
which may be solved in two manners: as a constrained minimisation, where nodes $\x^i$ and vertices $\by^I$ are independent and coupled through the constraint in \eqref{e:interp}, or by using this constraint in the expression of the objective function (total strain energy). We choose the latter approach in order to reduce the number of unknowns, and thus the size of the resulting system of equations. 

In order to deduce the expression of $\frac{\partial W_V}{\partial\x^i}$, we define first the vertex tractions as
\begin{align}\label{e:tv}
\bt^{IJ}_V=\frac{\partial W^{IJ}_V}{\partial\y^I}=-\bt^{JI}_V=-\frac{\partial W^{IJ}_V}{\partial\y^J}.
\end{align}

The nodal residuals due to contributions of the vertex network, denoted by $\g_V^i$, may be then computed by using the chain rule and the kinematic relation in \eqref{e:interp},
\begin{align}\label{e:g_V}
\g_V^i&:=\frac{\partial W_V}{\partial\x^i}
=\sum_{IJ}\left(\frac{\partial W^{IJ}_V}{\partial\y^I}\frac{\partial\y^I}{\partial\x^i}
+\frac{\partial W^{IJ}_V}{\partial\y^J}\frac{\partial\y^J}{\partial\x^i}\right)
=\sum_{IJ}\left(\bt^{IJ}_V p^i(\bxi^I) +\bt^{JI}_V p^i(\bxi^J)\right) \nonumber \\
& =\sum_{I\in B^i}p^i(\bxi^I)\sum_{J\in S^I}\bt^{IJ}_V.
\end{align}

In the last expression $B^i$ denotes the set of vertices that form the boundary of cell $i$, centred on $\x^i$, and $S^I$ is the set of vertices connected to vertex $I$. Note also that the last equality follows from the fact that $p^j(\bxi^K)$ vanishes if $K\notin B^j$. 

Total mechanical equilibrium is then found by solving the minimisation in \eqref{e:minix}, which yields,
\begin{align}\label{e:mecha}
\sum_{j\in S^i}\bt^{ij}_D + \sum_{I\in B^i}p^i(\bxi^I)\sum_{J\in S^I} \bt^{IJ}_V =\bO,\ i=1,\ldots,N_{nodes},
\end{align}
which in terms of the force contributions $\g^i_D$ and $\g^i_V$ reads
\begin{align}\label{e:gdgv}
\g^i_D + \g^i_V=\bO, \ i=1,\ldots,N_{nodes}.
\end{align}

The summation in the second term of \eqref{e:mecha} involves the vertex bars that have at least one vertex on the triangles that surround node $\x^i$. Figure \ref{f:nodesvertexB} shows a schematic view of how the boundary of each cell is defined within the tissue, and the traction vectors $\bt^{IJ}_V$ and $\bt^{JI}_V$.

Mechanical equilibrium of the system is obtained at cell centres (nodes) by solving the set of equations in 
\eqref{e:mecha}. Since this equation is non-linear with respect to the positions of the nodes, we resort to 
Newton-Raphson method for linearisation of the equations and to obtain the solution. The linearisation of the terms in \eqref{e:mecha} is given  in \ref{s:linearisation}.

Note that the second term in \eqref{e:mecha} arises due to the kinematic interpolation in \eqref{e:interp}. This term represents the nodal contribution of the vertex forces (reactions of the constraints in \eqref{e:interp}), which is proportional to the values of the shape functions $p^i(\bxi^I)$. This equation shows the coupling between nodal and vertex equilibrium. When vertex forces exist ($k_V\neq 0$), nodal forces and vertex forces are not necessarily equilibrated at nodes and vertex, respectively, that is, we may have that $\g_D^i\neq \bO$ and $\sum_{J\in S^I}\bt^{IJ}_V\neq \bO$. The latter condition is the equilibrium equation usually imposed in purely vertex models \cite{alt17}. We will analyse the evolution of these resultants in Section \ref{s:results} (Numerical results).

\subsection{Area constraint}

Cell volume invariance under tissue extension is relevant when the size and the number of
cells within the tissue is considered as constant. A two-dimensional area constraint will be imposed here by adding the energy term,
\begin{align}\label{e:W_A}
W_A=\frac{\lambda_A}{2}\sum_{m=1}^{\bar N_{nodes}}\left(A^m-A_0^m\right)^2,
\end{align}
where $\lambda_A$ is a penalisation coefficient and $A_0^m$ and $A^m$ are the initial and the current areas of cell $m$, respectively. The area of cell $m$ can be expressed in terms of its vertices by using Gauss theorem
\begin{align}\label{e:Am}
A^m=\int_{\Omega^m}dA=\frac{1}{2}\int_{\partial{\Omega^m}}\y\cdot\n ds,
\end{align}
where $\y$ is an arbitrary point on the boundary of cell $m$, $ds$ is the differential segment
of the cell boundary and $\n$ is the outward normal. Since each cell boundary forms a polygon, we will break the integral over the whole cell boundary into $N_m$ line integrals. Points between vertices $I$ and $J$ can be obtained by using a linear interpolation
\begin{align}\label{e:yq}
\y=q^I(\alpha)\y^I + q^J(\alpha)\y^J,
\end{align}    
with $\alpha\in\left[-1,1\right]$ a local coordinate along the cell boundary segment $IJ$, and  $q^I(\alpha)=\frac{1}{2}(1-\alpha)$ and $q^J(\alpha)=\frac{1}{2}(1+\alpha)$ the interpolation functions. By inserting equation \eqref{e:yq} into \eqref{e:Am} and noting that $ds=l^{IJ} d\alpha /2$, with $l^{IJ}=||\y^I-\y^J||$, we have
\begin{align}\label{e:Am2}
\begin{split}
A^m=
&\frac{1}{2}\sum_{IJ\in P^m}^{N_m}\int_{-1}^1\sum_I q^I(\alpha)\y^I\cdot \n^{IJ}\frac{l^{IJ}}{2}d\alpha
=\frac{1}{2}\sum_{IJ\in P^m}^{N_m}\frac{l^{IJ}}{2}\left(\y^I+\y^J\right)\cdot\n^{IJ},
\end{split}
\end{align}  
where $P^m$ denotes the segments of the polygon that surrounds node $\x^m$ (see Figure \ref{f:nodesvertexB}). The expression above can be simplified as
\begin{align}\label{e:Ai}
A^m
=\frac{1}{2}\sum_{\substack{IJ\in P^m}}^{N_m}\left(\y^I\times\y^J\right)\cdot\e_z
=\frac{1}{2}\sum_{\substack{IJ\in P^m}}^{N_m}\y^I\cdot\J\y^J
\end{align}
with $\J=\left[\begin{array}{cc} 0&-1\\1&0\end{array}\right]=-\J^T$ and such that $(\by^I\times\by^J)\cdot\e_z=\by^I\cdot\J\by^J$. Finally, the total area of the whole set of $\bar N_{nodes}$ cells in the tissue, $A_T$, can be expressed as
\begin{align}\label{e:A_T}
A_T=\frac{1}{2}\sum_{m=1}^{\bar N_{nodes}}\sum_{IJ\in P^m}\y^I\cdot\J\y^J.
\end{align} 

The expression of the contribution in \eqref{e:Ai} is inserted in the energy term in \eqref{e:W_A}, and appended to the total elastic energy, 
\[
W=W_D(\x)+W_V(\by(\x))+W_A(\by(\x)),
\]
which is minimised with respect to the nodal positions $\x^i$. This gives rise  to an additional nodal contribution, 
\begin{align}\label{e:g_A}
\g^i_A&:=\frac{\partial W_A}{\partial \x^i}= 
\frac{\lambda_A}{2}\J\sum_{m\in \bar S^i}\left(A^m-A^m_0\right)\sum_{IJ\in P ^m}\left(p^i(\bxi^I)\y^J-p^i(\bxi^J)\y^I\right).
\end{align}

The set $\bar S^i$ in the first summation includes the nodes that surround node $i$ and also node $i$ itself. Since the force vector above is non-linear, the Jacobian must be complemented with additional terms arising from the linearisation of $\g^i_A$. These terms are given in \ref{s:linearisation}.

\subsection{$\xi$-Relaxation}\label{s:xirelax}

When the values of $\bxi^I$ are kept constant, vertices and cell-centred positions are coupled through the constraint in \eqref{e:interp}. As pointed out in Section \ref{s:mvertex}, this constraint has the effect of altering the usual equilibrium conditions in cell-centred and vertex networks (vanishing of the sum of forces at nodes and at vertices, respectively). In fact, in our equilibrium equations in \eqref{e:mecha} and \eqref{e:gdgv}, the additional force due to $\g_V^i$ (which contains the tractions $\bt^{IJ}_V$) may be regarded as a reaction force stemming from the constraints in \eqref{e:interp}.  This modified equilibrium may furnish non-smooth and unrealistic deformations at the tissue boundaries, which can then exhibit a zig-zag shape.

In order to avoid these effects, we will disregard the constraint \eqref{e:interp} for those vertices at the boundary, and relax the value of $\bxi^I$, which can attain values different from $\{1\ \ 1\}^T/3$. Those vertices are then allowed to change their relative positions within the corresponding  triangle $\mathcal T^I$, and may be not necessarily located at the barycentre. In this case, mechanical equilibrium is expressed as a vanishing sum of tractions at the vertex location, as it is customary in vertex models \cite{okuda15,perrone16}. In our hybrid model, we interpret the parametric coordinates $\bxi$ of those vertices as additional unknowns. The energy terms including the vertices are now made dependent on these extra parametric coordinates, i.e. we write $W_V(\y(\x, \bxi))$ and $W_A(\y(\x, \bxi))$.

When relaxing the constraint, we will further limit the increments of $\bxi$ between time-steps, so that their positions are kept not too far from their otherwise  interpolated value in order to minimise large discontinuities between discrete time-points on the resulting force contributions. This is achieved by adding to the total energy of the system $W$ and at each time $t_{n+1}$ a term that penalises the variations of $\bxi$,
\begin{align}\label{e:Wxi}
W_\xi(\bxi)&=\frac{\lambda_\xi}{2} \sum_{I\  relaxed}||\bxi_{n+1}^I-\bxi_n^I||^2.
\end{align}

By interpreting the factor $\lambda_\xi$ as a viscous coefficient $\approx \frac{\eta}{\Delta t}$, this additional term is equivalent to a viscous-like effect, since it generates forces proportional to the incremental vertex positions (or vertex velocities).

The extension of the system with additional variables $\bxi$ also modifies the minimisation problem in \eqref{e:minix}, which now takes the form
\begin{align}\label{e:minixi}
\{\x^*, \bxi^*\}=\argmin_{\x, \bxi} W(\x, \bxi),
\end{align}
with
\begin{align}\label{e:W}
W(\x, \bxi)=W_D(\x) + W_V(\by(\x,\bxi)) + W_A(\by(\x,\bxi))+W_\xi(\bxi).
\end{align}

Equilibrium is now represented by two systems of equations,
\begin{align}\label{e:g}
\vect g
&:=\left\{
\begin{array}{c}
\g_x\\
\g_y
\end{array}
\right\}=\bO,
\end{align}
with $\g_x=\nabla_{\x} W(\x, \bxi)$ and $\g_y=\nabla_{\bxi} W(\x, \bxi)$. Each residual contribution in the total residual $\g$ is the sum of different energy contributions in \eqref{e:W}, so that $\g=\g_D+\g_V+\g_A+\g_\xi$, where each term  contains in turn nodal ($\x$) and vertex ($\bxi$) contributions, 
\begin{align}\label{e:gxgy}
\begin{aligned}
\g^i_x&:=\frac{\partial W(\x, \bxi)}{\partial \x^i}=\g^i_D+\g_V^i+\g_A^i+\g^i_\xi,\\
\g^I_y&:=\frac{\partial W(\x, \bxi)}{\partial \bxi^I}=\g^I_D+\g_V^I+\g_A^I+\g^I_\xi.
\end{aligned}
\end{align}

Since the nodal strain energy $W_D$ does not depend on $\bxi^I$, and the penalty term $W_\xi$ does not depend on the nodal positions $\x^i$ (see equations \eqref{e:W_D} and \eqref{e:Wxi}), we have that  $\g_D^I=\bO$ and $\g_\xi^i=\bO$. The nodal contributions $\g^i_D$, $\g^i_V$ and $\g_A^i$ have been given respectively in \eqref{e:g_D}, \eqref{e:g_V} and \eqref{e:g_A}. The vertex  contributions require the computations of
\begin{align}\label{e:dydx}
\nabla_\bxi W &=
\nabla_\bxi W_V +\nabla_\bxi W_A+\lambda_\xi(\bxi_{n+1}-\bxi_n)\nonumber \\
\frac{\partial \y^I}{\partial \bxi^I}&=\sum_{\x^i\in \mathcal T^I}\x^i\otimes \nabla p^i(\bxi^I)
\end{align}
so that we have, also from equations \eqref{e:W_A} and \eqref{e:Ai}, 
\begin{align}\label{e:g_xi}
\g^I_V &:=\frac{\partial W_V}{\partial\bxi^I}
=\sum_{JK}\frac{\partial W^{JK}_V}{\partial\y^J}\frac{\partial\y^J}{\partial\bxi^I}
+\frac{\partial W^{JK}_V}{\partial\y^K}\frac{\partial\y^K}{\partial\bxi^I}
=\sum_{K\in S^I} \sum_{\x^i\in \mathcal T^I}(\bt^{IK}_V\cdot\x^i)\nabla p^i(\bxi^I)\nonumber \\
\g^I_A &:=\lambda_A\sum_{m=1}^{\bar N_{nodes}}(A^m-A^m_0)\frac{\partial A^m}{\partial\bxi^I}\\
\g^I_\xi&:=\nabla_\bxi W_\xi =\lambda_\xi(\bxi_{n+1}^I-\bxi_n^I) \nonumber
\end{align}
with $\frac{\partial \y^I}{\partial \bxi^I}$ given in \eqref{e:dydx}, and 
\begin{align}\label{e:dAI}
\frac{\partial A^m}{\partial\bxi^I}=\frac{1}{2}\sum_{KL\in P^m}\left(
\delta_{KI}\left(\frac{\partial \y^K}{\partial \bxi^I}\right)^T\J\y^L
-\delta_{LI}\left(\frac{\partial \y^L}{\partial \bxi^I}\right)^T\J\y^K
\right).
\end{align}
 
The symbol $\delta_{KI}$ above is the Kronecker delta, which is equal to $1$ if $K=I$ and $0$ otherwise. We note that if we extended $\xi$-relaxation to the whole tissue, we could recover standard vertex models, that is, a model where the vertices positions are solely determined by their mechanical equilibrium: sum of forces at each vertex equal to zero. In our numerical simulations we have though just applied $\xi$-relaxation to specific boundaries of the domain.

\section{Rheological model}\label{s:rheo}

So far, the bar elements of the cell-centred and vertex networks have been considered as purely elastic, with a strain function given in equations \eqref{e:W_D} and \eqref{e:W_V} respectively. Since cells exhibit both elastic and viscous response \cite{harris12}, we here extend the elastic strain energy function of the bars with the ability to vary their resting length $L$. The rate of change  of the resting length is given by the evolution law
\begin{align}\label{e:ldot}
\frac{\dot L}{L}=\gamma\veps
\end{align}
where $\gamma$ is the \emph{remodelling rate}, and $\veps$ is the elastic strain used either in \eqref{e:W_D} or \eqref{e:W_V}. It has been previously shown that such a rheological model is equivalent to a Maxwell viscoelastic behaviour \cite{munoz13b}, and that can be used to simulate tissue fluidisation  \cite{asadipour16} or cell cortex response \cite{doubrovinsky17}.

In order to include the inherent contractility that cells exert \cite{salbreux12b}, the previous evolution law is modified as
\begin{align}\label{e:evol}
\frac{\dot L}{L}=\gamma(\veps-\veps^c)
\end{align}
with $\veps^c$ a contractility parameter. This modification aims to attain a homoeostatic elastic strain equal to $\veps^c$,  at which no further modifications of the resting length take place. 

The ordinary differential equation (ODE) in \eqref{e:evol} is employed for the bar elements of the nodal and vertex networks, and it is solved together with the non-linear equations in \eqref{e:minixi}. In fact, the evolution law is taken into account by first discretising in time the ODE in \eqref{e:evol} with a $\beta$-weighted scheme. By using the strain definition $\veps=(l-L)/L$, the discretisation of \eqref{e:evol}  yields
\begin{align}\label{e:devol}
L_{n+1}-L_n=\Delta t\gamma(l_{n+\beta}-L_{n+\beta}-\veps^cL_{n+\beta}),
\end{align}
with $(\bullet)_{n+\beta}=(1-\beta)(\bullet)_n+\beta(\bullet)_{n+1}$. In our numerical tests we have used the value $\beta=0.5$. The discretisation in \eqref{e:devol} allows us to write
\begin{align}\label{e:dLdl}
\frac{\partial L}{\partial l}=\frac{\beta\Delta t\gamma}{1+\beta\Delta t\gamma(1+\veps^c)}.
\end{align}

This term is inserted in the traction definitions of $\bt^{ij}_D$ ad $\bt_V^{IJ}$ in \eqref{e:td} and \eqref{e:tv}, which are then computed with the help of the following derivation,
\begin{align}\label{e:deps}
\frac{\partial \veps^{ij}}{\partial \x^i}=\frac{1}{L}\left(1-\frac{l}{L}\frac{\partial L}{\partial l}\right)\e^{ij}\quad, \quad
\frac{\partial \veps^{IJ}}{\partial \y^I}=\frac{1}{L}\left(1-\frac{l}{L}\frac{\partial L}{\partial l}\right)\e^{IJ},
\end{align}
with
\begin{align}\label{e:eij}
\e^{ij}=-\e^{ji}=\frac{\x^i-\x^j}{||\x^i-\x^j||}\quad , \quad 
\e^{IJ}=-\e^{JI}=\frac{\y^I-\y^J}{||\x^i-\x^j||}.
\end{align}

The traction forces in \eqref{e:td} and \eqref{e:tv} read then respectively,
\begin{align}\label{e:tdtv}
\begin{aligned}
\bt_D^{ij}&=\frac{\partial W_D^{ij}}{\partial\x^i}
=\frac{\veps^{ij}}{L^{ij}}\left(1-\frac{l^{ij}}{L^{ij}}\frac{\partial L^{ij}}{\partial l^{ij}}\right)\e^{ij},\\
\bt_V^{IJ}&=\frac{\partial W_V^{IJ}}{\partial\y^I}
=\frac{\veps^{IJ}}{L^{IJ}}\left(1-\frac{l^{IJ}}{L^{IJ}}\frac{\partial L^{IJ}}{\partial l^{IJ}}\right)\e^{IJ}.
\end{aligned}
\end{align}
 
\section{Remodelling: Equilibrium-Preserving Map}\label{s:map}

One of the key features of soft biological tissues is their ability to remodel, that is, to change their neighbouring cells during growth, mobility and morphogenesis. We aim to include this feature in our model by computing a new connectivity $\T_{n+1}$ after each time point $t_n$. In this work we resort to the Delaunay triangulation of the nodal network, which guarantees a minimum aspect-ratio of the resulting triangles. We also assume that these optimal aspect ratios will not be exceedingly spoiled during tissue deformation. 

\begin{figure}[!htb]
\centering
\includegraphics[width=1.0\textwidth]{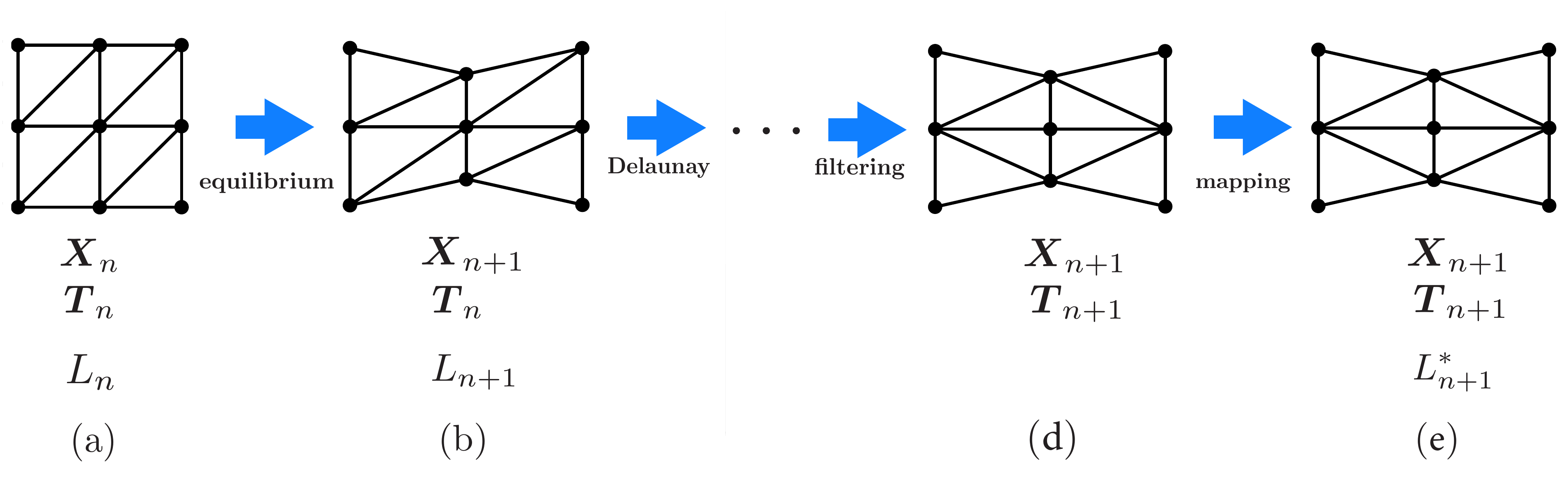}
\caption{ Deformation and remodelling process, including the computation of the resting lengths $L_{n+1}^*$ through the Equilibrium-Preserving Map, which maintains the network connectivity and nodal and vertex positions.}
\label{f:EPM}
\end{figure}

The redefinition of the network topology from $\T_n$ to $\T_{n+1}$ may involve drastic changes in the nodal and vertex equilibrium equations. Furthermore, the resting lengths $L^{ij}$ and $L^{IJ}$ are undefined for the newly created bar elements. In order to smooth mechanical transition between time-steps, we will here present an \emph{Equilibrium-Preserving Map} that computes $L^{ij}$ and $L^{IJ}$ by minimising the error of the mechanical equilibrium for the new connectivity. We will consider two approaches: a map that preserves the nodal and vertex equilibrium in a coupled manner (\emph{full-network mapping}), and a map that preserves nodal equilibrium and vertex equilibrium independently (\emph{split-network mapping}). The computational process depicted in Figure \ref{f:steps} is now  completed with the EPM as shown in Figure \ref{f:EPM}.

\subsection{Full-network mapping}\label{s:full}

In this approach, we aim to compute a new set of resting lengths $L^{ij}$ and $L^{IJ}$ that minimises the functional
\begin{align}\label{e:pi}
\hat \pi_F(L^{ij},L^{IJ})=\sum_i^{nodes}\left\|\gt_D^i+\gt_V^i+\g_{A}^i-\br^i\right\|^2.
\end{align}

This functional measures the error in the mechanical equilibrium considering all the residual contributions at node $i$ due to the cell-centres ($\gt_D^i$), the vertex network ($\gt_V^i$) and area constraints ($\g_{A}^i$).  The latter is the value obtained from the expression in \eqref{e:g_A}, while $\br^i$ is the total reaction for those nodes that have prescribed displacements. The residual contributions are computed as a function of nodal and vertex tractions as
\begin{align}\label{e:tpi}
\begin{aligned}
\gt_D^i&=\sum_{j\in S^i} \btt_D^{ij}=\sum_{j\in S^i} k_D\left(\frac{l^{ij}}{L^{ij}}-1\right)\e^{ij}\\
\gt_V^i&=\sum_{I\in B^i}p^i(\bxi^I)\sum_{J\in S^I} \btt_V^{IJ}
=\sum_{I\in B^i}p^i(\bxi^I)\sum_{J\in S^I} k_V\left(\frac{l^{IJ}}{L^{IJ}}-1\right)\e^{IJ}
\end{aligned}
\end{align}

Note that $\btt_D^{ij}$ are $\btt_V^{IJ}$ are not defined as $\frac{\partial W_D^{ij}}{\partial \x^{i}}$ or $\frac{\partial W_V^{IJ}}{\partial \y^{I}}$, but with a simpler purely elastic law, which disregards any rheological  evolution of the resting lengths. 

We emphasise that while computing the new resting lengths and thus the variables $L^{ij}$ and $L^{IJ}$, the nodal and vertex positions $\x^i$ and $\y^I$, and also the current lengths $l^{ij}$ and $l^{IJ}$, are all constant.

The minimisation of $\hat \pi_F$ in \eqref{e:pi} gives rise to a non-linear system of equations in terms of the unknowns $L^{ij}$ and $L^{IJ}$, but that is linear with respect to the inverse of these quantities. We will denote these inverses by $\theta^{ij}=1/L^{ij}$ and $\theta^{IJ}=1/L^{IJ}$. The new functional, denoted by $\pi_F(\theta^{ij}, \theta^{IJ})$, is obtained by inserting this change of variables 
\begin{align*}
(\theta^{ij},\theta^{IJ})^*=\argmin \pi_F(\theta^{ij},\theta^{IJ}).
\end{align*}
 
The optimal variables $\theta^{ij}{}^*$ and $\theta^{IJ}{}^*$ are found by solving the associated normal equations of this least-squares problem, which after making use of \eqref{e:tpi} reads
\begin{align}\label{e:Mma}
\left[\begin{array}{cc}
\A_{DD} & \A_{DV} \\
\A_{DV}^T & \A_{VV} \\
\end{array}\right]\left\{\begin{array}{c}
\btheta_D\\
\btheta_V
\end{array}\right\}=\left\{\begin{array}{c} 
\bb_D\\
\bb_V
\end{array}\right\}
\end{align}
with $\btheta_D$ and $\btheta_V$ vectors containing all the inverses of the resting lengths for the nodal and vertex networks, $1/L^{ij}$ and $1/L^{IJ}$ respectively, and 
\begin{align}\label{e:A}
\begin{aligned}
\A^{mn,pq}_{DD}=&k_D^2l^{mn}{\e^{mn}}^T\left(\sum_{j\in S^m}l^{mj}\e^{mj}\delta_{mj}^{pq}-\sum_{j\in S^n}l^{nj}\e^{nj}\delta_{nj}^{pq} \right)\\
\A_{DV}^{mn,PQ}=&k_D k_Vl^{mn}{\e^{mn}}^T\left(
\sum_{I\in B^m}p^m(\bxi^I)\sum_{J\in S^I} l^{IJ}\e^{IJ}\delta_{IJ}^{PQ}\right.\\
&\left.-\sum_{I\in B^n}p^n(\bxi^I)\sum_{J\in S^I} l^{IJ}\e^{IJ}\delta_{IJ}^{PQ}
\right) \\
\A_{VV}^{MN,PQ}=&k_V^2\sum_i^{N_{nodes}}
\left(\sum_{I\in B^i} p^i(\bxi^I)\sum_{J\in S^I}l^{IJ}\e^{IJ}\delta_{IJ}^{PQ}\right)\\
& \left(\sum_{I \in B^i}p^i(\bxi^I)\sum_{J\in S^I}l^{IJ}\e^{IJ}\delta^{MN}_{IJ}\right)\\
\bb_{D}^{mn}=&k_Dl^{mn}\left({\hat\g}^m-{\hat \g}^n\right)^T\e^{mn}\\
\bb_{V}^{MN}=&\sum_i^{N_{nodes}}k_V{\hat \g}^i{}^T
\left(\sum_{I\in B^i}p^i(\bxi^I)\sum_{J\in S^I}l^{IJ}\e^{IJ}\delta^{MN}_{IJ}\right)
\end{aligned}
\end{align}

In the equations above, we have defined
\begin{align}\label{e:ghat}
\begin{aligned}
{\hat \g}^i&= k_D\sum_{j\in S^i}l^{ij}\e^{ij}+k_V\sum_{I\in B^i} p^i(\bxi^I)\sum_{J\in S^I}\e^{IJ}-\g_A^i+\br^i\\
\delta_{mj}^{pq}&=\left\{\begin{array}{r l}
1,& \ \mbox{if}\  mj=pq, \mbox{or}\ mj=qp, \\
0, & \ \mbox{otherwise.}
\end{array}\right.\\
\delta_{IJ}^{PQ}&=\left\{\begin{array}{r l}
1,& \ \mbox{if}\  IJ=PQ, \mbox{or}\ IJ=QP, \\
0, & \ \mbox{otherwise.}
\end{array}\right.
\end{aligned}
\end{align}

The uniqueness of the solution of system of equations in \eqref{e:Mma}, and thus the regularity of the system matrix, is in general not guaranteed, since more than one combination of tractions in equilibrium with the reaction field may be found in some cases. This is algebraically reflected by a large condition number of the system matrix. For this reason, the functional is regularised by adding an extra term,
\begin{align}\label{e:pir}
\pi_{F\lambda}(\theta^{ij}, \theta^{IJ})=\pi_{F}(\theta^{ij}, \theta^{IJ})
+\lambda_L\left(\sum_{ij}
||\theta^{ij}-\frac{1}{l^{ij}}||^2
+\sum_{IJ}||\theta^{IJ}-\frac{1}{l^{IJ}}||^2\right)
\end{align}
with $l^{ij}$ and $l^{IJ}$ the current distances between connected nodes and vertices, respectively. This regularisation adds a factor $\lambda_L$ on the diagonal components and factors $\lambda_L/l^{mn}$ and  $\lambda_L/l^{MN}$ on $\bb_D^{mn}$ and $\bb_V^{MN}$, which ensure that the system will have a unique solution for a sufficiently large value of the regularisation parameter $\lambda_L$. In our numerical examples we have used $\lambda_L=10^{-12}$.

\subsection{Split-network mapping}

The previous approach allows to find equilibrated tractions with a possible redistribution of forces between the vertex and nodal networks. In some cases though, it is desirable to keep the traction contributions of the two networks split. For this reason, we present an alternative Equilibrium-Preserving Map that aims to compute the resting lengths by considering equilibrium conditions for the nodal and vertex networks independently. This is achieved by minimising the functional
\begin{align}\label{e:piS}
\pi_S(\theta^{ij},\theta^{IJ})=\pi_D(\theta^{ij}) + \pi_V(\theta^{IJ})
\end{align}
with
\begin{align*}
\pi_D(\theta^{ij}) &=\sum_{i}^{N_{nodes}}||\tilde\g^i_D-\br_D^i||^2  \\
\pi_V(\theta^{IJ}) &=\sum_{i}^{N_{nodes}}||\tilde\g^i_V-\br_V^i||^2 
\end{align*}
where $\br_D^i$ is the contribution from the nodal network on node $i$ before remodelling, and $\br_V^i$ is the contribution from the vertex network to node $i$ before remodelling. This contributions are obtained from the residual contributions before remodelling takes place as
\begin{align}\label{e:r}
\begin{aligned}
\br_D^i&=\g_D^i,\\
\br_V^i&=\g_V^i+\g_A^i.
\end{aligned}
\end{align}

Applying the same approach as in Section \ref{s:full} to $\pi_F$, the minimisation of $\pi_S$ yields two uncoupled systems of equations,
\begin{align}
\begin{split}
&\A_{DD}\btheta_D=\bb'_D\\
&\A_{VV}\btheta_V=\bb'_V.
\end{split}
\end{align}

Matrices $\A_{DD}$ and $\A_{VV}$ are those written in equation \eqref{e:A}, while the right-hand-sides are now given by
\begin{align*}
\bb_{D}'{}^{mn}=&k_Dl^{mn}\left({\hat\g}^m_D-{\hat \g}^n_D\right)^T\e^{mn}\\
\bb_{V}'{}^{MN}=&\sum_i^{N_{nodes}}k_V{\hat \g}^i_V{}^T
\left(\sum_{I\in B^i}p^i(\bxi^I)\sum_{J\in S^I}l^{IJ}\e^{IJ}\delta^{MN}_{IJ}\right)
\end{align*}
with
\begin{align*}
{\hat \g}^i_D&= k_D\sum_{j\in S^i}l^{ij}\e^{ij}+\br^i_D,\\
{\hat \g}^i_V&= k_V\sum_{I\in B^i} p^i(\bxi^I)\sum_{J\in S^I}\e^{IJ}+\br^i_V.
\end{align*}

Like in the previous section, a regularisation term, equal to the one used in \eqref{e:pir}  is added to the functional $\pi_S$ in order to ensure the regularity and uniqueness of the solution, with the same value of the regularisation parameter $\lambda_L=10^{-12}$.

The split-network approach is in fact relevant when the stresses in the nodal and vertex networks follow different patterns, and it is necessary  to maintain this difference between the networks, such as wound healing, where the stresses around the wound ring are significantly higher. Preserving stress residuals independently at each network guarantees the stress contrast. The full-network approach on the other hand, spoils this contrast and may transfer some of the stresses on the wound ring to the nodal network. The numerical example in Section \ref{s:wound} illustrates this fact.

\section{Numerical results}\label{s:results}

\subsection{Extension of square tissue}

We test our methodology by extending a square domain obtained from a random perturbation of a $10\times10$ grid of nodes (see Figure \ref{f:extension}a). The domain is formed by 81 cells, and subjected to a uniform 30\% extension applied within 60 time-steps. We will test two situations: extension with constant topology (evolution from (a)-(b)), and with remodelling (evolution (a)-(c)). In the two situations we will apply the full and split approaches of the Equilibrium-Preserving Map (EPM).
\begin{figure}[!htb]
\centering
\includegraphics[scale=.4]{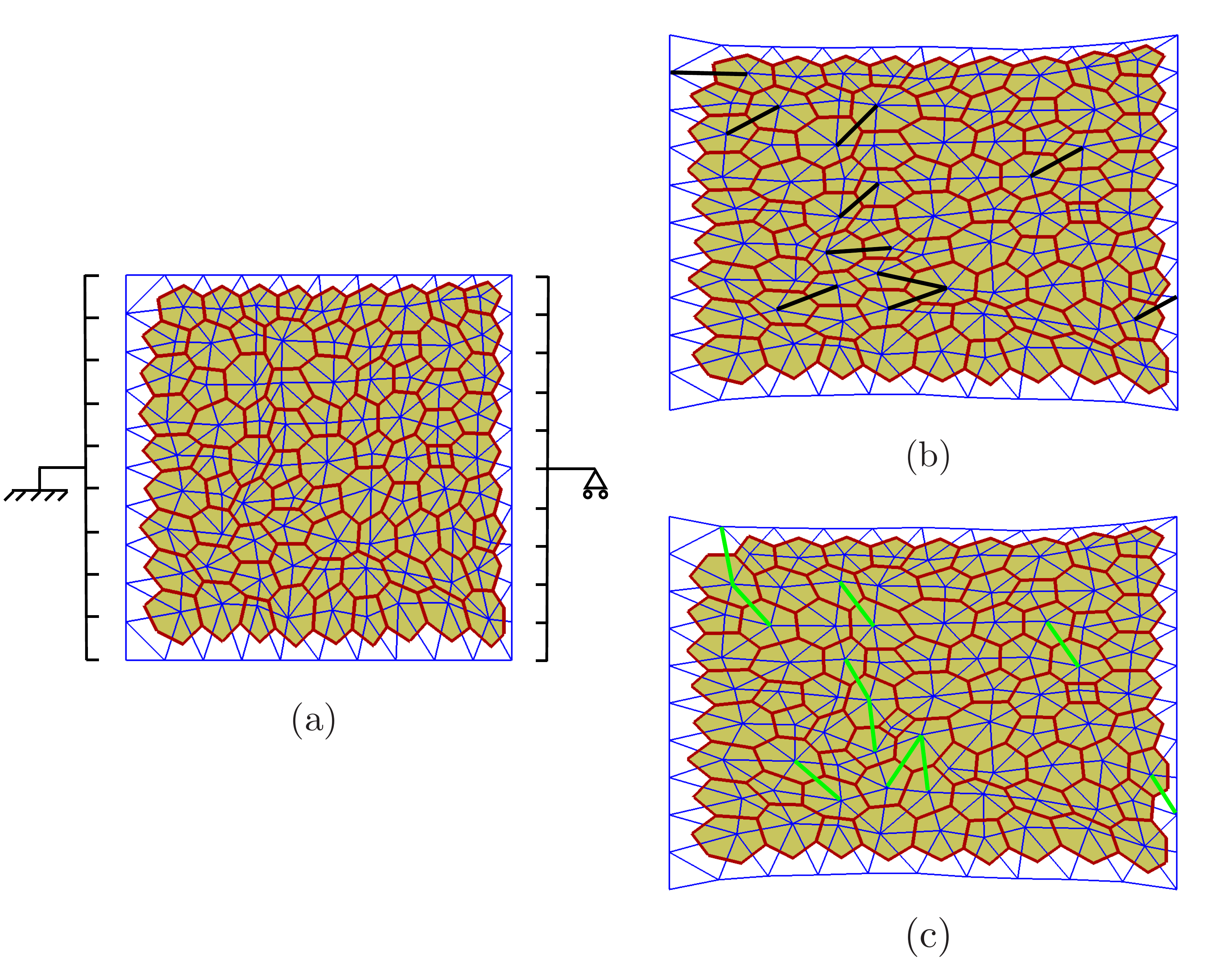}
\caption{Tissue extension. (a) Initial configuration, (b) tissue configuration at 30\% extension without remodelling, and (c) tissue configuration at 30\% extension with remodelling. Replaced elements are marked in black in (b). Remodelled elements are marked in green.}
\label{f:extension}
\end{figure}
%


\subsubsection{Validation of EPM: fixed topology}\label{map_test}

To inquire the accuracy and effects of the EPM, we measure the total reaction at the right side and the elastic energy of the tissue during extension while keeping the topology constant. Figure \ref{f:map} shows the evolution of the two quantities when  $k_D=0.1 k_V$ (Figures \ref{f:map}a-b) and when  $k_D=10 k_V$ (Figures \ref{f:map}c-d). It can be observed that in all cases the full-network and the split-network mappings give the same values as the tests with no mapping. This fact shows that the EPM is able to recover the same traction values as the ones when no computation of the resting lengths is applied, and that the system regularisation is not altering these lengths or the elastic response of the tissue.

\begin{figure}[!htb]
\subcaptionbox{\label{map_DstiffR}}
 {
 \hspace{-5ex}
 \includegraphics[width=.58\linewidth]{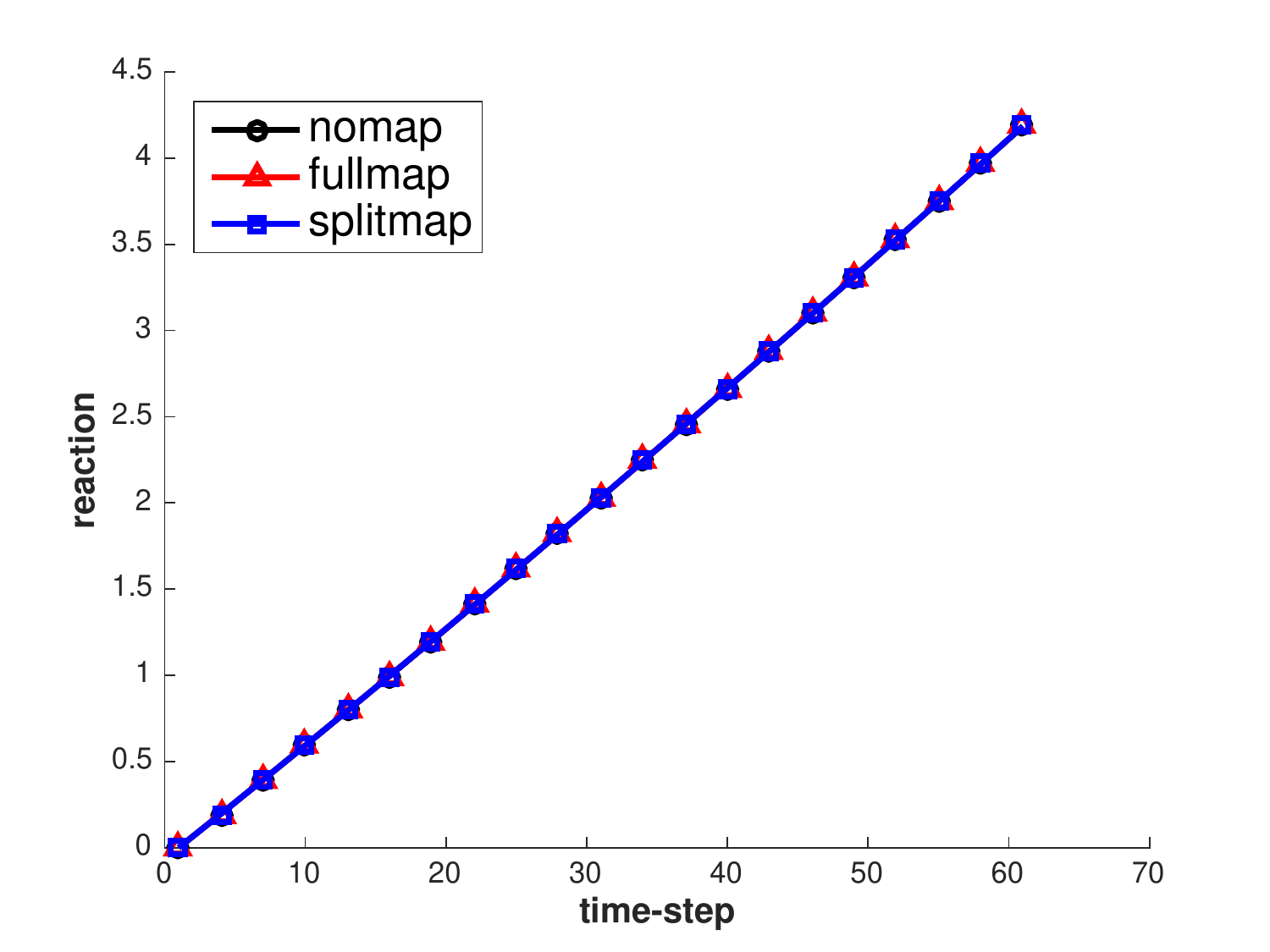}
 }
  \hspace{-11ex}
 \subcaptionbox{\label{map_DstiffE}}{
 \includegraphics[width=.58\linewidth]{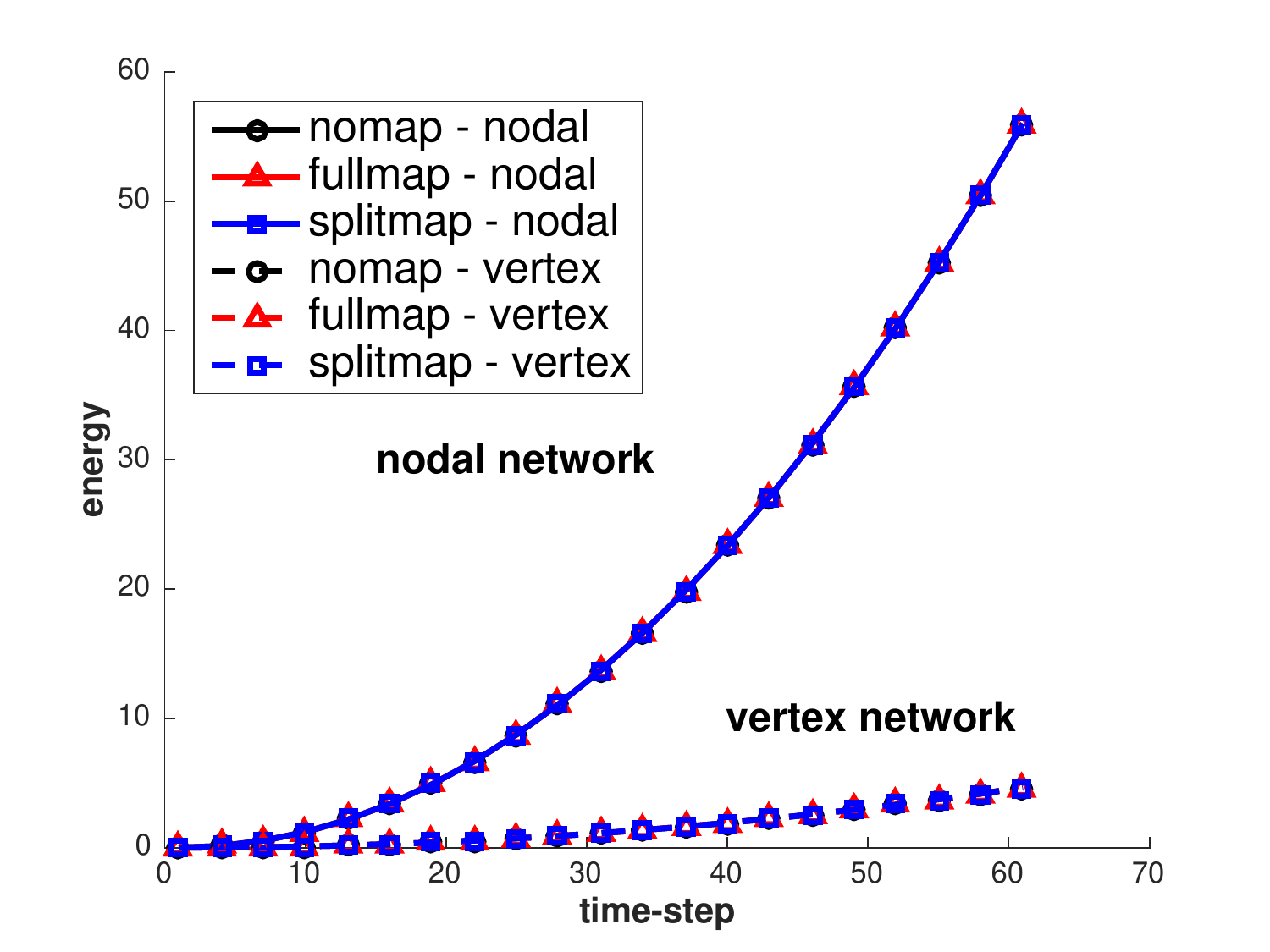}
} 
\ContinuedFloat
\subcaptionbox{\label{map_VstiffR}}{
 \centering
 \hspace{-5ex}
 \includegraphics[width=.58\linewidth]{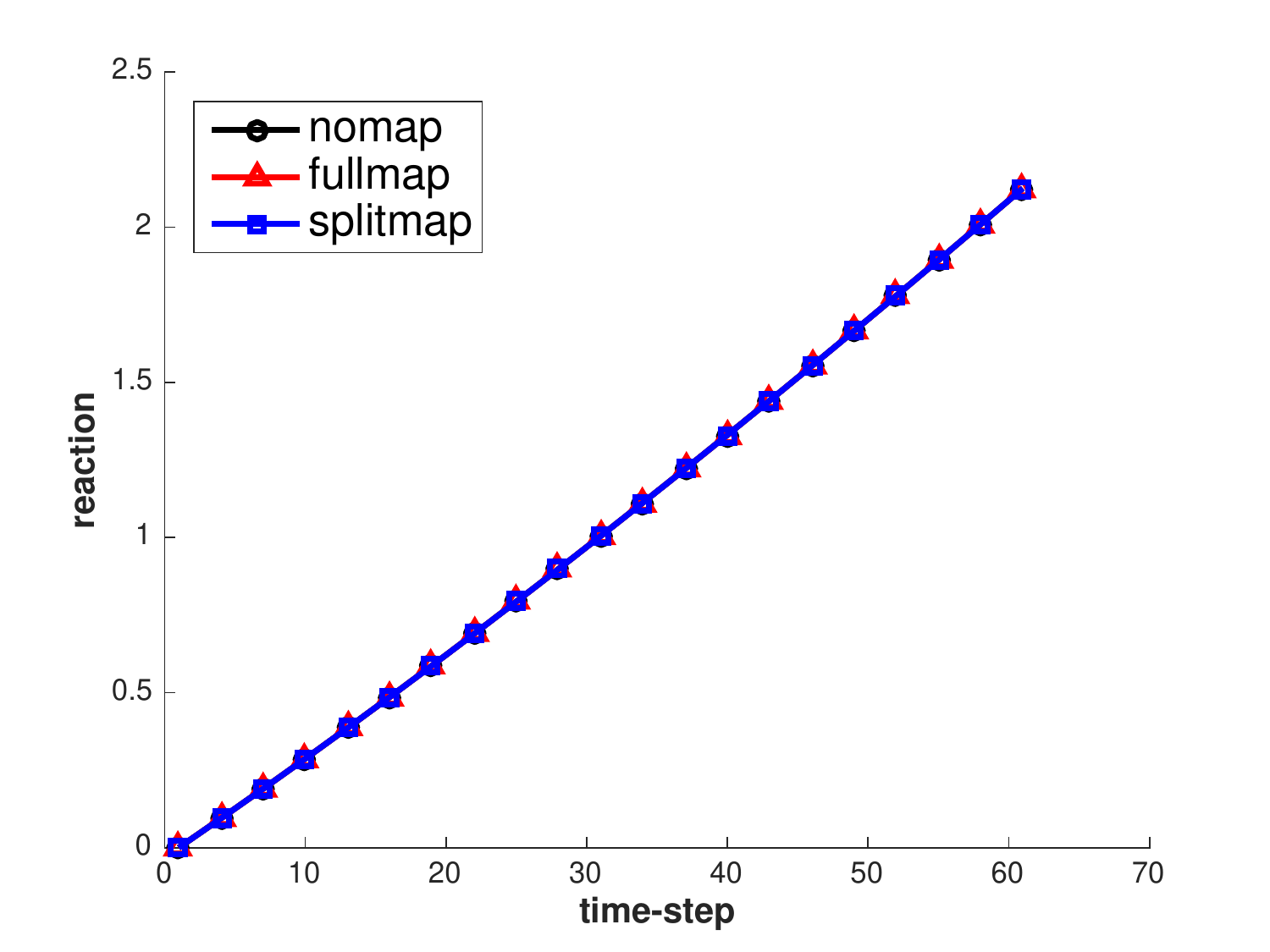}
}
\hspace{-11ex}
\subcaptionbox{\label{map_VstiffE}}{
\centering
\includegraphics[width=.58\linewidth]{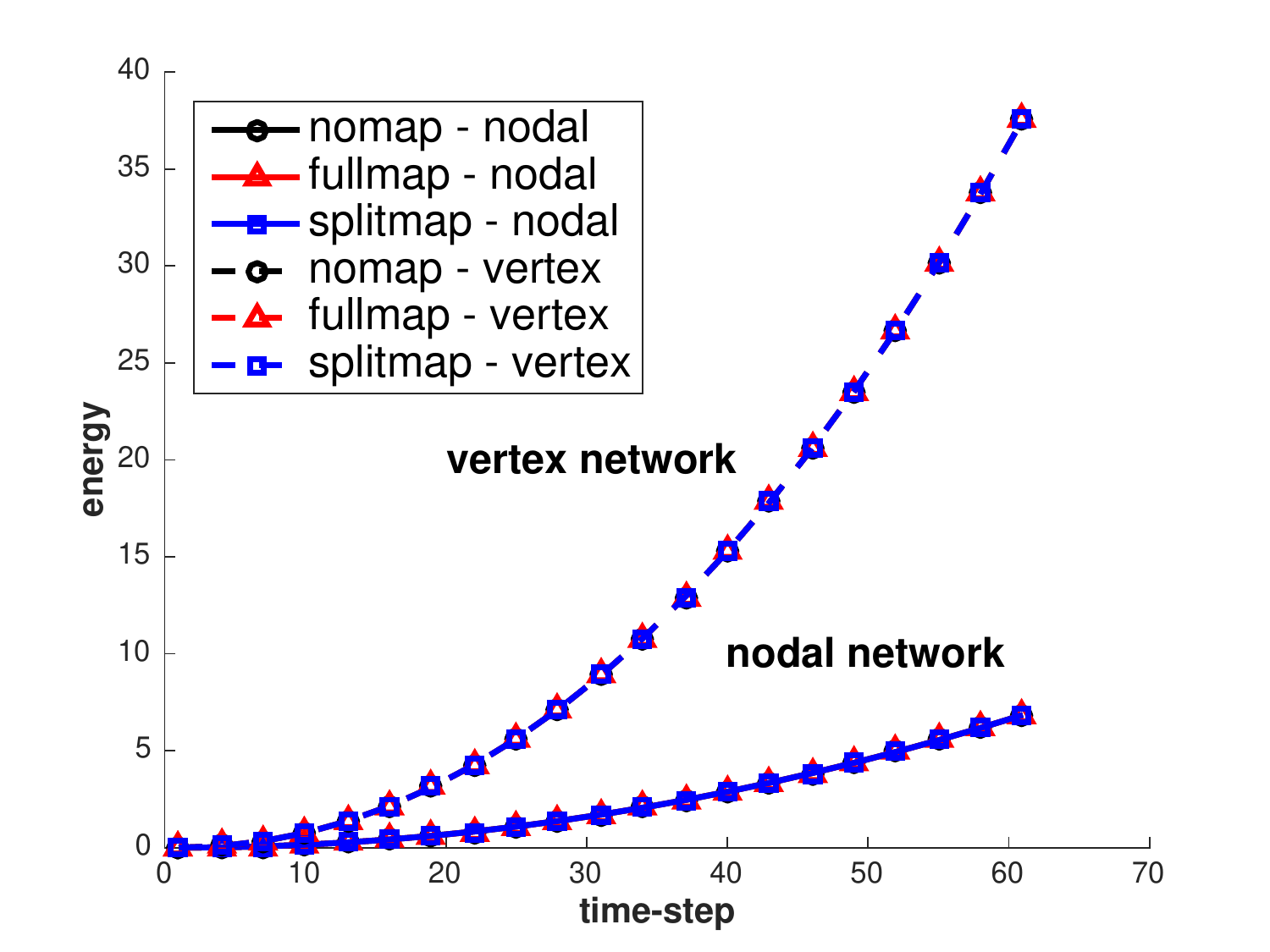}
}
\caption{ Tissue formed by linear elastic elements, under 30\% uniform stretch applied within 60 time-steps while held at constant topology (no remodelling). Elements resting lengths, at each time-step, obtained by three approaches: fixed resting lengths (no network mapping), full-network mapping and split-network mapping. (a) Total tissue reaction while $k_D=10 k_V$. (b) Potential energy of nodal and vertex networks while $k_D=10 k_V$. (c) Total tissue reaction while $k_D=0.1 k_V$. (d) Potential energy of nodal and vertex networks while $k_D=0.1 k_V$.}
\label{f:map}
\end{figure}


\subsubsection{Validation of the EPM: variable topology}

We now apply the same boundary conditions as in the previous tests, but allowing the tissue to remodel according to the Delaunay triangulation of the nodal positions. Figure \ref{f:rem} shows the total reaction at the right end and the total elastic energy. We have monitored these quantities under three conditions: no remodelling/mapping, remodelling with full-network mapping and remodelling with split-network mapping. We have tested also two sets of material properties: $k_D=10k_V$ (Figure \ref{f:rem}a-b), and $k_D=0.1k_V$ (Figure \ref{f:rem}c-d). The total  number of remodelling events (elements that change their connectivity) is also plotted at each time-step, whenever this number is positive.   

\begin{figure}[!htb]
\subcaptionbox{\label{rem_DstiffR}} {
\hspace{-5ex}
\includegraphics[width=.58\linewidth]{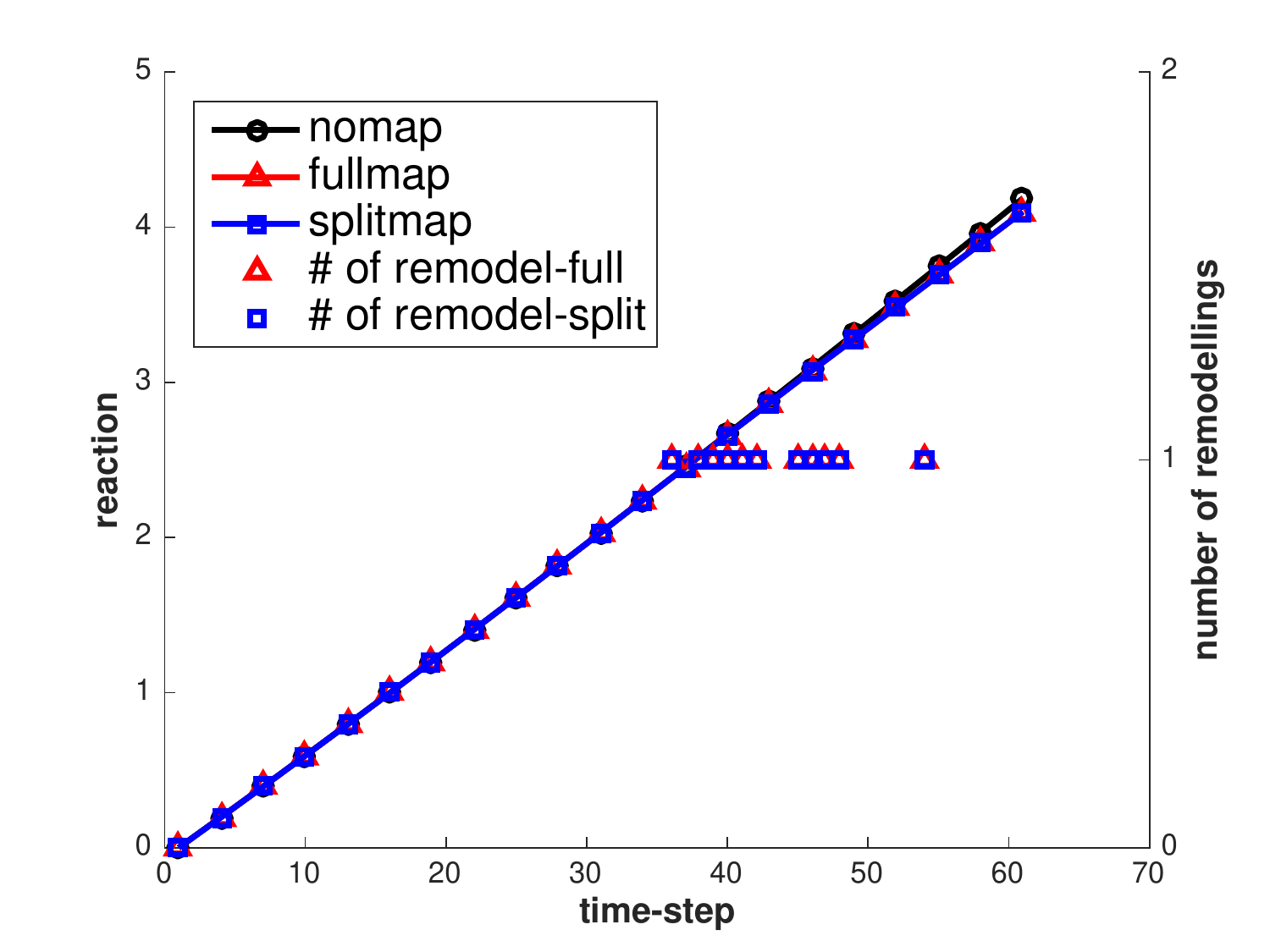}
 }
\hspace{-11ex}
 \subcaptionbox{\label{rem_DstiffE}}{
 \includegraphics[width=.58\linewidth]{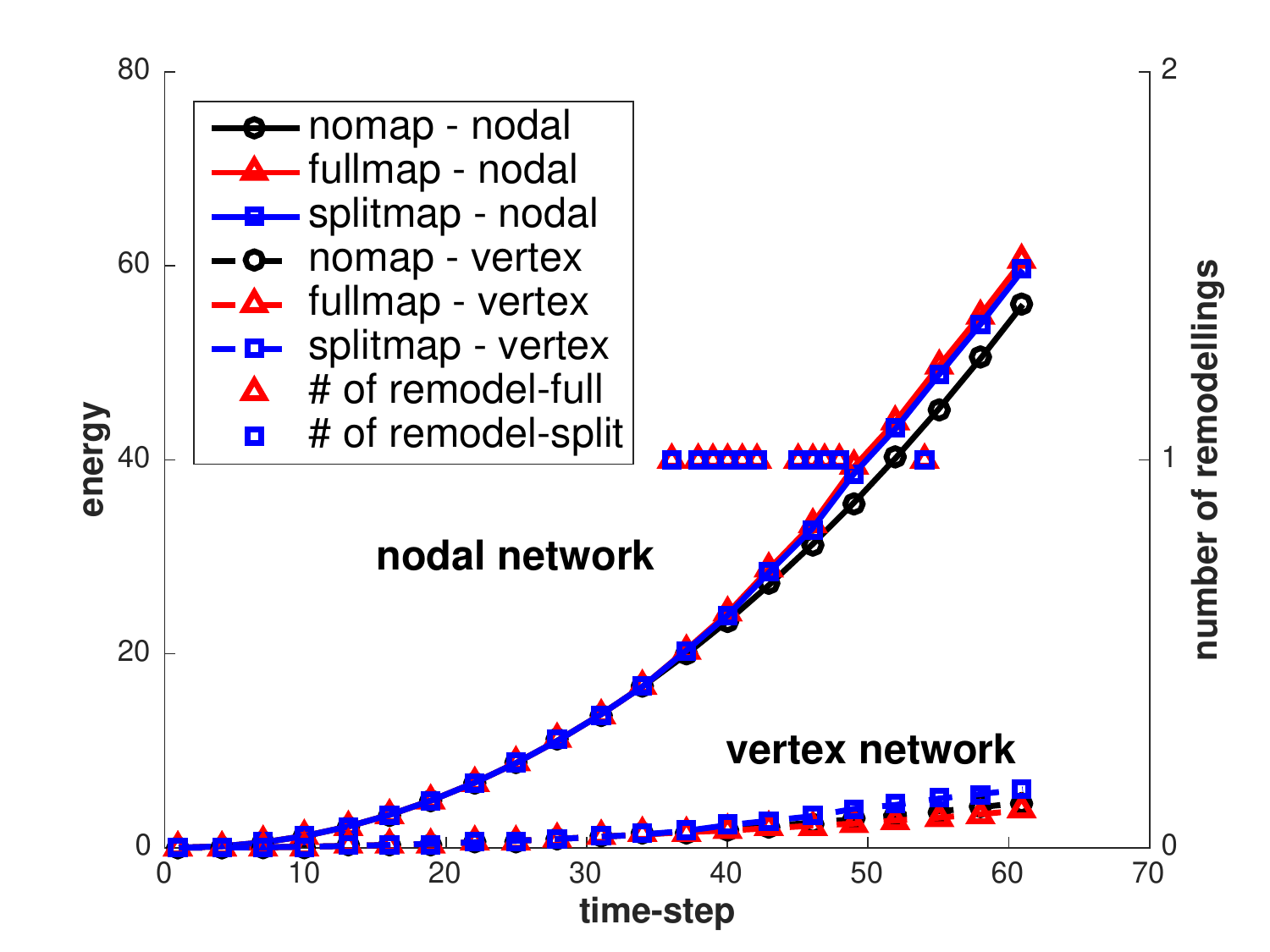}
} 
\ContinuedFloat
\subcaptionbox{\label{rem_VstiffR}}{
 \centering
\hspace{-5ex}
 \includegraphics[width=.58\linewidth]{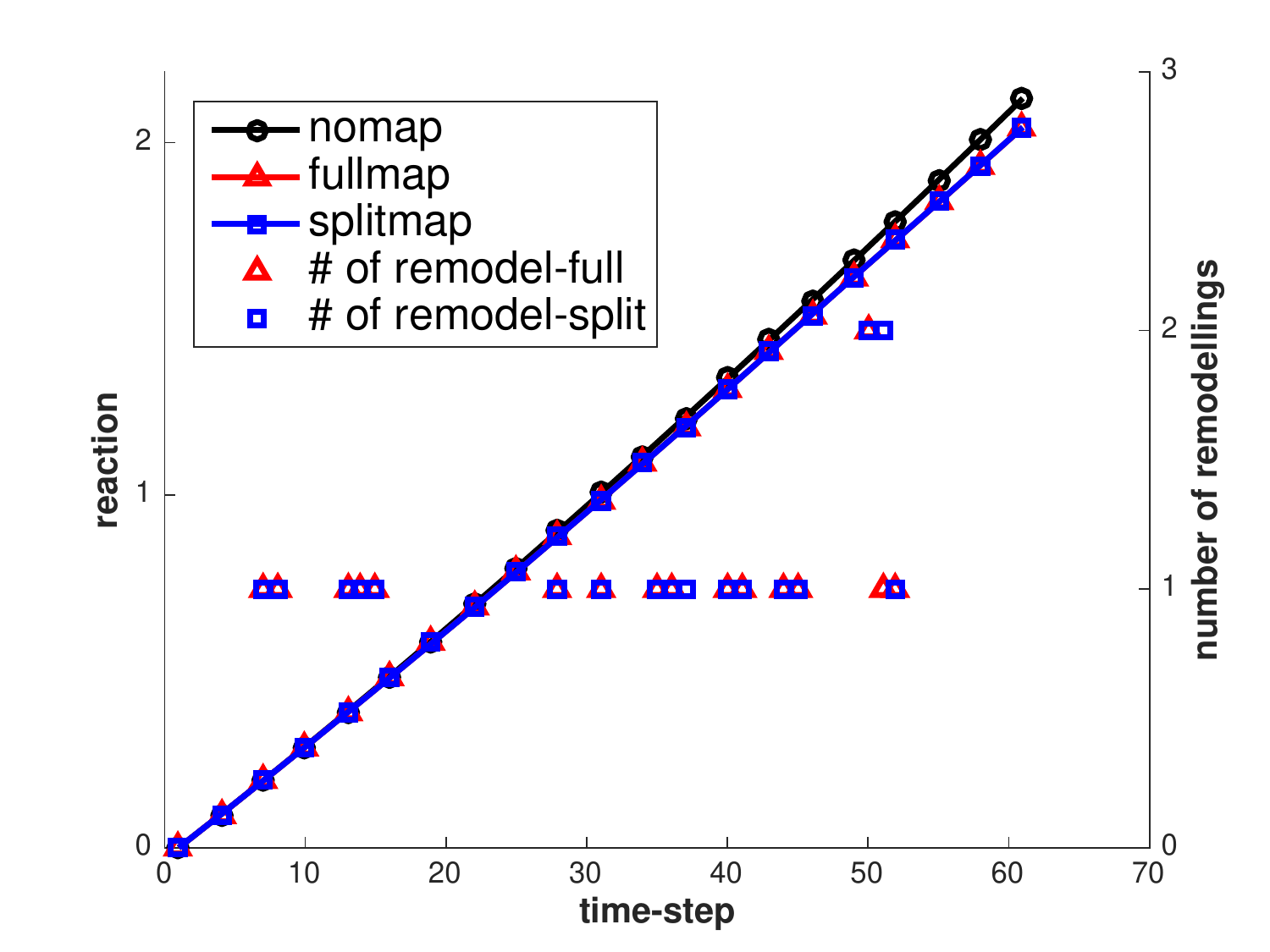}
}
\hspace{-11ex}
\subcaptionbox{\label{rem_VstiffE}}{
\centering
\includegraphics[width=.58\linewidth]{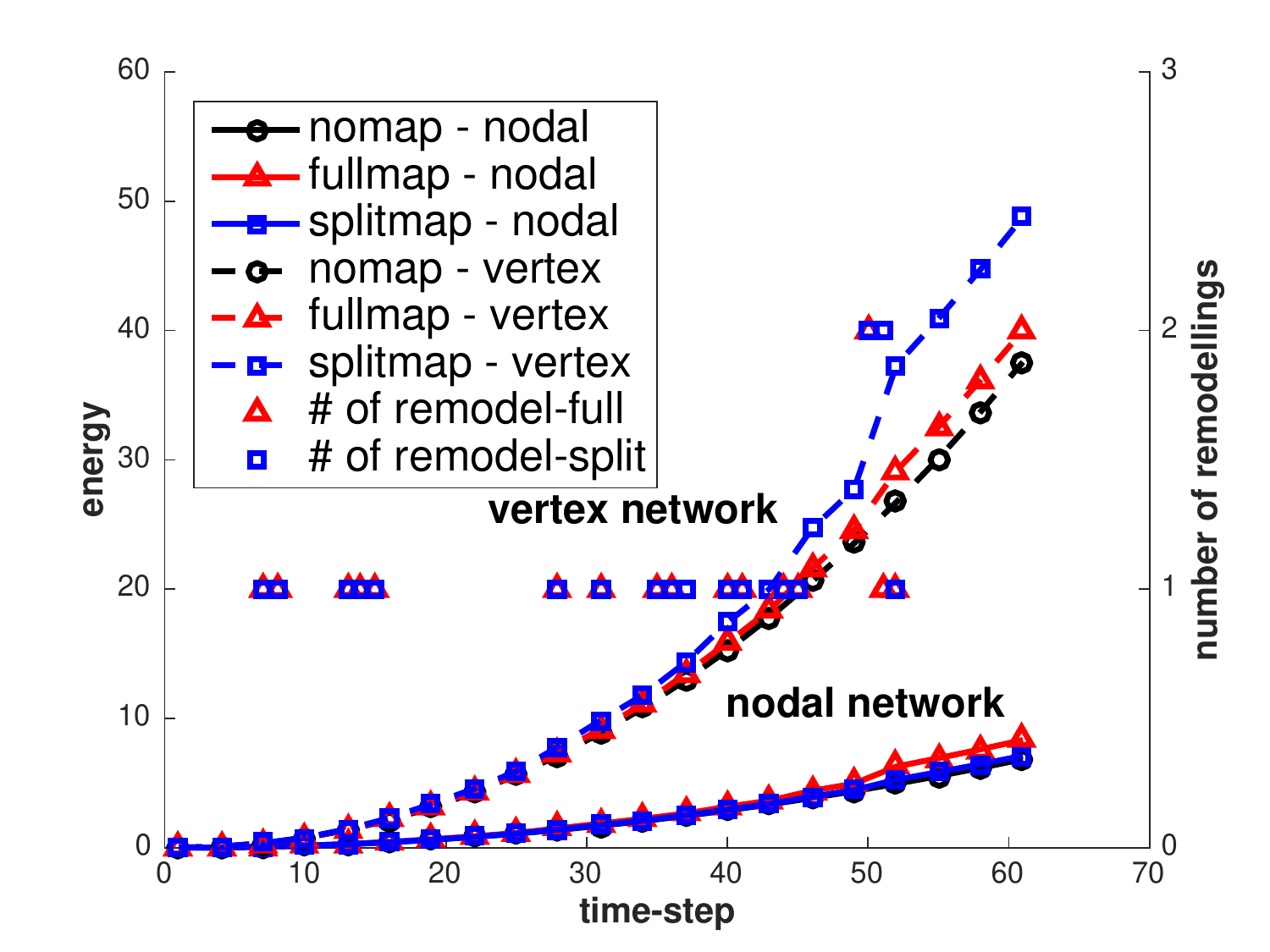}
}
\caption{ Tissue formed by linear elastic elements, under 30\% uniform stretch applied within 60 time-steps which is allowed to remodel. Elements resting lengths, at each time-step, obtained by three approaches: fixed resting lengths (no network mapping/remodelling), full-network mapping and split-network mapping with floating topology. (a) Total tissue reaction while $k_D=10 k_V$. (b) Elastic energy of nodal and vertex networks while $k_D=10 k_V$. (c) Total tissue reaction while $k_D=0.1 k_V$. (d) Elastic energy of nodal and vertex networks while $k_D=0.1 k_V$.}
\label{f:rem}
\end{figure}

From the plots in Figure \ref{f:rem} it can be observed that the evolution of the total reaction is not substantially affected by the remodelling process. The elastic energy, however, suffers some deviations with respect to the case with no remodelling when the split-network EPM is used and the vertex network is stiffer than the nodal network. This drift is more severe when more remodelling events are encountered. Indeed, the split-network approach prevents the transfer of energy between the vertex and nodal networks, preventing in some cases the full preservation of the equilibrium conditions before the remodelling events. The total reaction of the tissue is in all cases not much affected by the mapping, which is in agreement with the fact that EPM aims to compute resting lengths distributions that match the nodal resultants before remodelling. 

For the two sets of material parameters, the total reaction, and thus the tissue response, is very much unaffected by the remodelling  for the two EPM approaches. This allows to keep the correct aspect ratio of the cells while keeping the elastic response. Although cells may use remodelling events to relax their stress state, we here aim to independently control the stress relaxation and the remodelling events. In our example, the stress relaxation is prevented by using a small value of the remodelling rate $\gamma=10^{-6}$. 

\subsubsection{Analysis of $\xi$-relaxation}

Tissue stiffness against tissue total reaction and strain energy is investigated by assigning a range of values to $\{k_D\ k_V\}$ at a constant total stiffness, $k_D+k_V=1$, under two conditions: 1) when vertices are  rigidly anchored at barycentres ($\bxi=\frac{1}{3}\{1\ \ 1\}$), and 2) when vertices are allowed to change their relative positions with respect to the barycentres ($\bxi$-relaxation).  Figure \ref{f:xi_compare} compares the vertex network shown in Figure \ref{f:extension}b for the two situations. The red network displays vertices anchored at barycenters, while in the green network vertices are relaxed under a penalisation factor $\lambda_\bxi=10^{-4}$.  
\begin{figure}[!htb]
\centering
\includegraphics[scale=0.3]{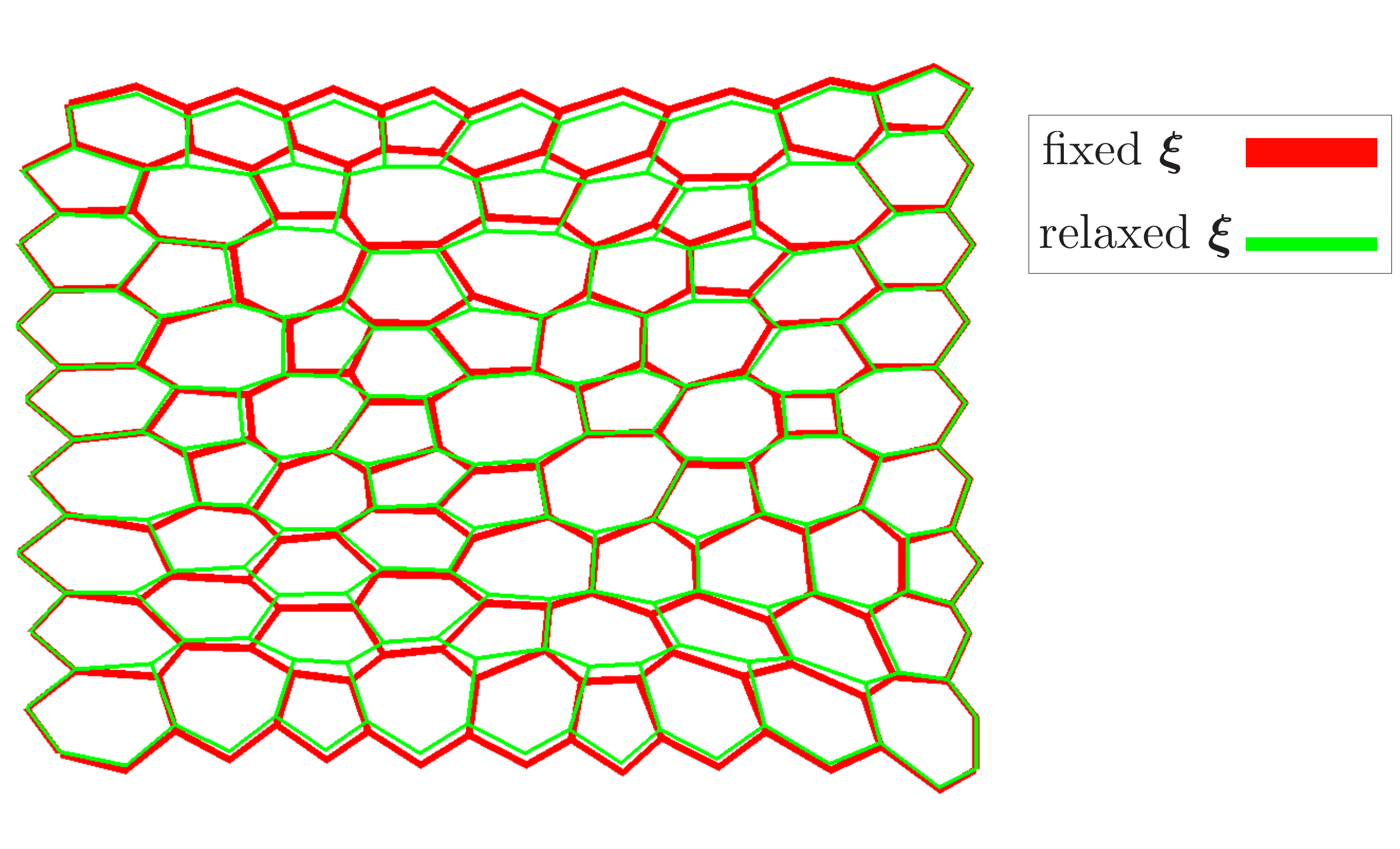}
\caption{Deformed tissue at 30\% extension. Red network represents vertices with fixed $\bxi$. Green network represents vertices when $\bxi$-relaxation is  allowed.}
\label{f:xi_compare}
\end{figure}

In order to inspect the effect of $\xi$-relaxation we have analysed the reaction and energy of mainly nodal-driven or mainly vertex-driven tissues for different values of $\lambda_\xi$. Figure \ref{f:xi_plot} shows the tissue response for different values of $k_V\in[0,1]$ while keeping $k_V+k_D=1$, and when the tissue is subjected to an 30\% extension.  Figure \ref{f:xi_plot}a shows that the total reaction decreases as  tractions concentrate on the vertex network. This reduction is steeper when vertices are relaxed (lower values of $\lambda_\xi$). Figure \ref{f:xi_plot}b shows a faster drop in tissue total energy  and a lower growth in vertex network energy, while no significant effect on nodal network energy when $\bxi$-relaxation is allowed. 

We have also analysed the difference of our equilibrated tractions with respect to the purely nodal and vertex equilibrium conditions: null sum of tractions at nodes and at vertices. This difference is computed as the mean value of the following nodal and vertex measures,
\begin{align}\label{e:Ei}
\begin{aligned}
E_i&=\frac{||\sum_{j\in S^i}\vect t^{ij}_D||}{\sum_{j\in S^i}||\vect t^{ij}_D||}, i=1,\ldots,N_{nodes}\\
E_I&=\frac{||\sum_{J\in S^I}\vect t^{IJ}_V||}{\sum_{J\in S^I}||\vect t^{IJ}_V||}, I=1,\ldots,N_{tri}
\end{aligned}
\end{align}

Figures \ref{f:xi_plot}c and \ref{f:xi_plot}d plot the means $\bar E_D=\sum_i E_i/N_{nodes}$ and $\bar E_V=\sum_I E_I/N_{tri}$ for the whole tissue.  As expected, the nodal difference is zero when no stiffness is assigned to the vertex network ($k_V=0$). As $k_V$ increases, pure nodal equilibrium is increasingly violated, due to the coupling between the two networks. In most cases, this difference is below $10\%$, except when vertices are fixed. Pure vertex equilibrium is more severely affected by the kinematic constraint, but the difference also decreases rapidly as $\lambda_\xi$ decreases. It can be observed that while the positions of the vertices in the two networks is very similar, purely vertex equilibrium drastically improves for approximately $\lambda_\xi<10^{-2}$.

\begin{figure}[hbt]
\subcaptionbox{\label{xi_R}} {
\hspace{-4ex}
\includegraphics[width=.53\linewidth]{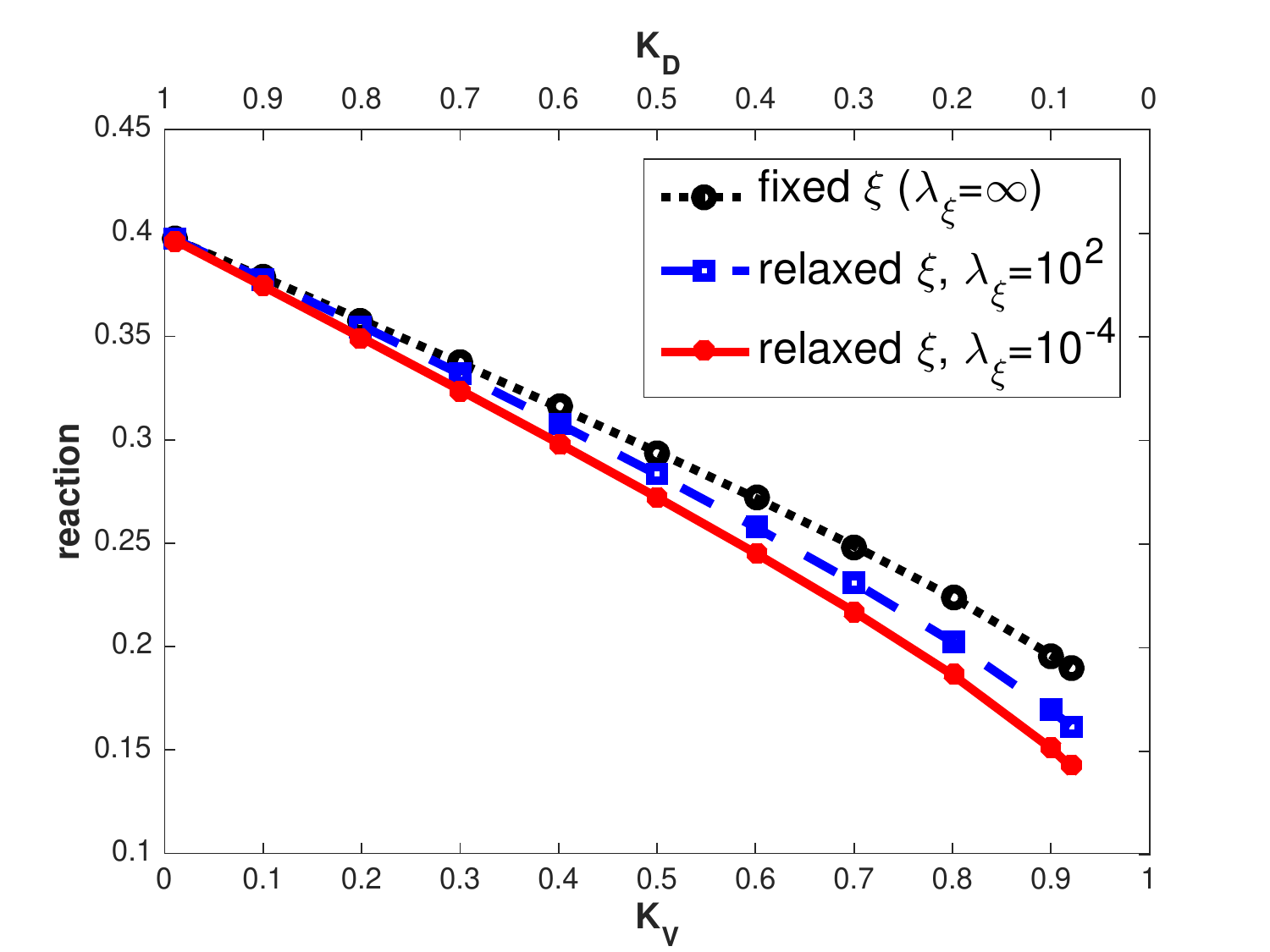}
 }
\hspace{-4ex}
 \subcaptionbox{\label{xi_E}}{
 \includegraphics[width=.53\linewidth]{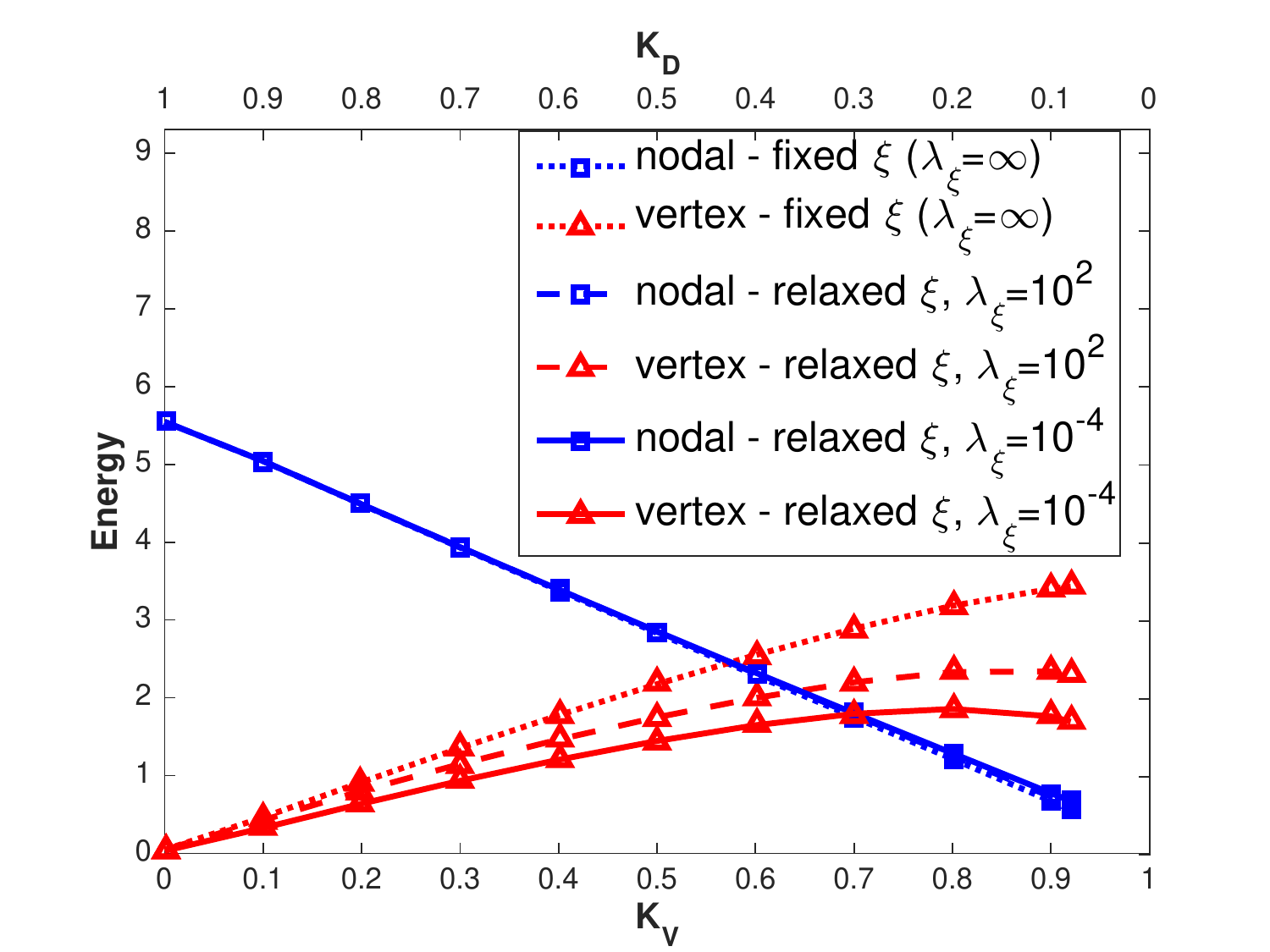}
} 
\ContinuedFloat
\subcaptionbox{\label{xi_erD}}{
 \centering
 \hspace{-4ex} 
 \includegraphics[width=.53\linewidth]{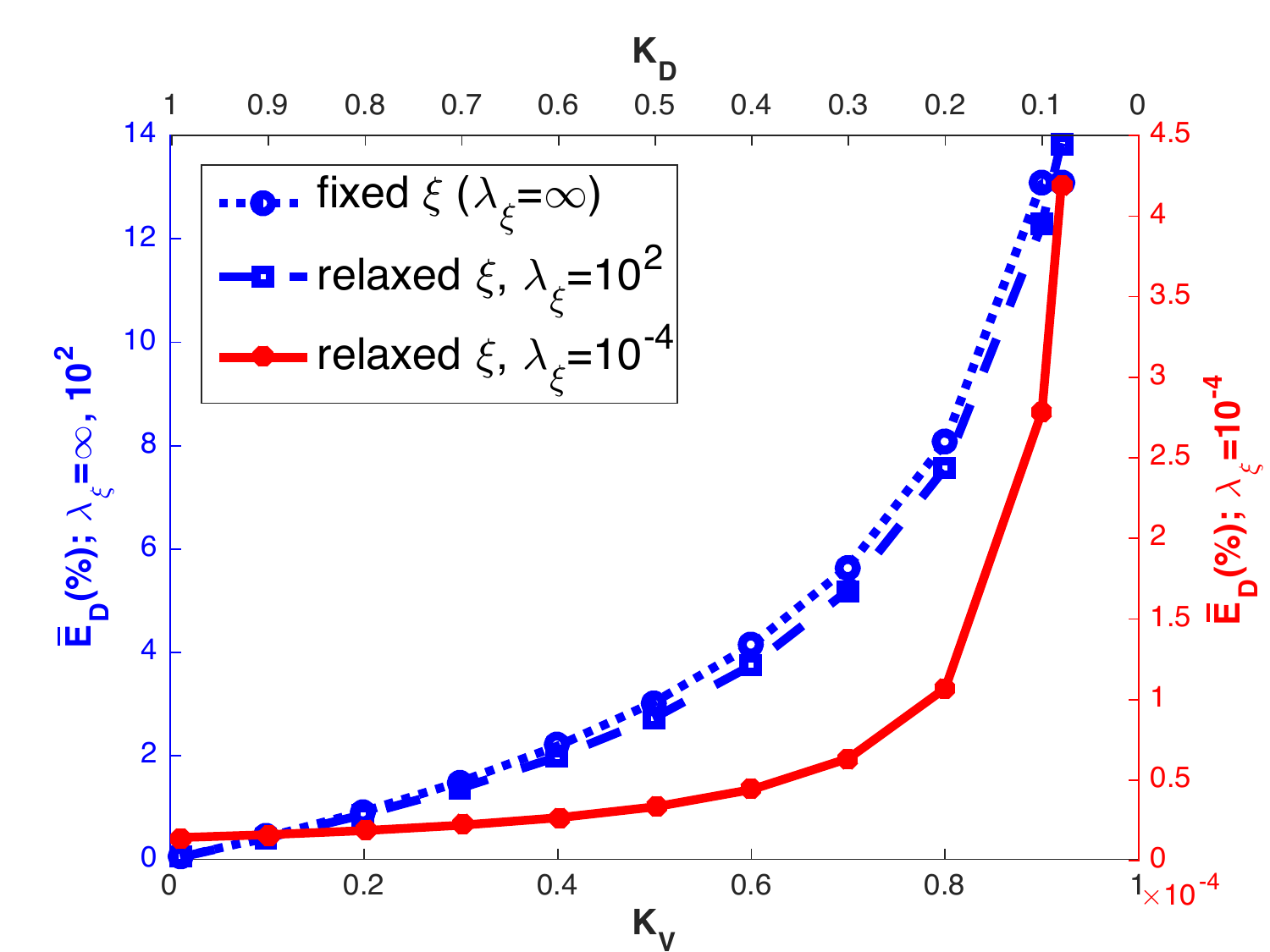}
}
 \hspace{-4ex} 
\subcaptionbox{\label{xi_erV}}{
\centering
\includegraphics[width=.53\linewidth]{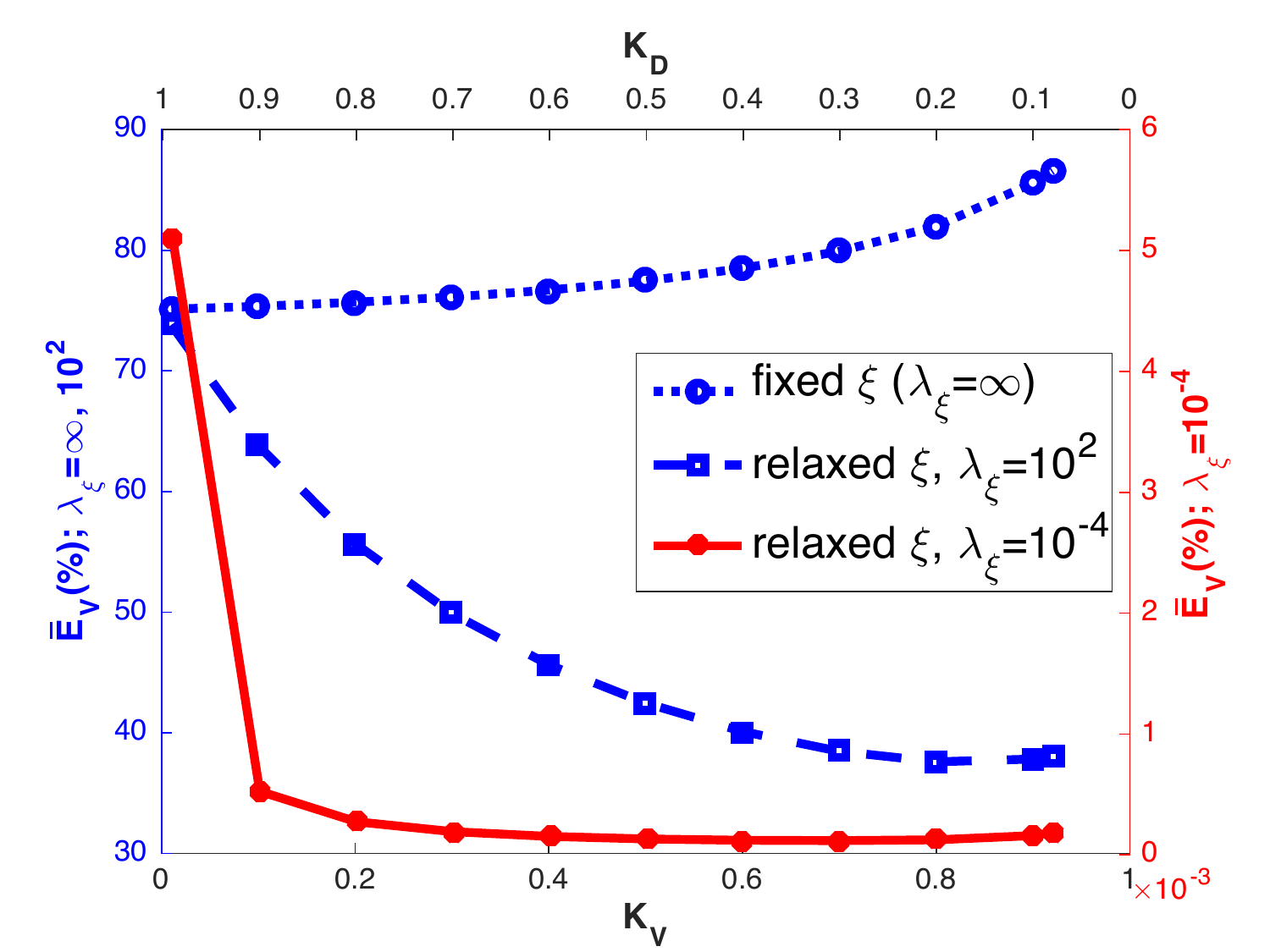}
}
\caption{ Analysis of response of tissue composed of elastic elements, under 30\% uniform stretch applied within a single time-step for different combinations of $\{k_D\ k_V\}$ while $k_D+k_V=1$, with and without $\bxi$-relaxation. (a) Tissue reaction. (b) Nodal, vertex and total strain energy of the tissue. (c) Mean of the difference between pure nodal and coupled equilibrium $\bar E_D$ for different values of $\lambda_\xi$. (d) Mean of the difference between coupled and pure vertex equilibrium $\bar E_V$ for different values of $\lambda_\xi$ (note the difference on the scaling of the left and right vertical axes). See equation in \eqref{e:Ei} and text below for the definitions of $\bar E_V$ and $\bar E_V$.}
\label{f:xi_plot}
\end{figure}

\subsection{Wound healing}\label{s:wound}

The model is tested  to simulate a wound healing process in monolayers \cite{antunes13}. The evolution law in \eqref{e:evol} is applied to the nodal and vertex networks with the values given in Table \ref{t:par}, which also indicates  that the area constraint is imposed in order to mimic mechanical properties of the tissue. Topological changes in the tissue are allowed to examine the role of cell motility and cell intercalation during wound healing.

\begin{table}[!htb]
\begin{center}
\begin{tabular}{|l|l|| l| l|| l| l|| l|}
\hline
$k_D$ & $k_V$ & $\gamma_D$ & $\gamma_V$ & $\veps^c_D$ & $\veps^c_V$ & $\lambda_A$ \\
\hline
0.1 & 2.0 & 0.5 & 0.5 & 1.0 & 0.7 & 10.0 \\
\hline
\end{tabular}
\end{center}
\caption{Material parameters employed in the wound healing example.}
\label{t:par}
\end{table}

\begin{figure}[!htb]  
\centering
\includegraphics[width=1.0\textwidth]{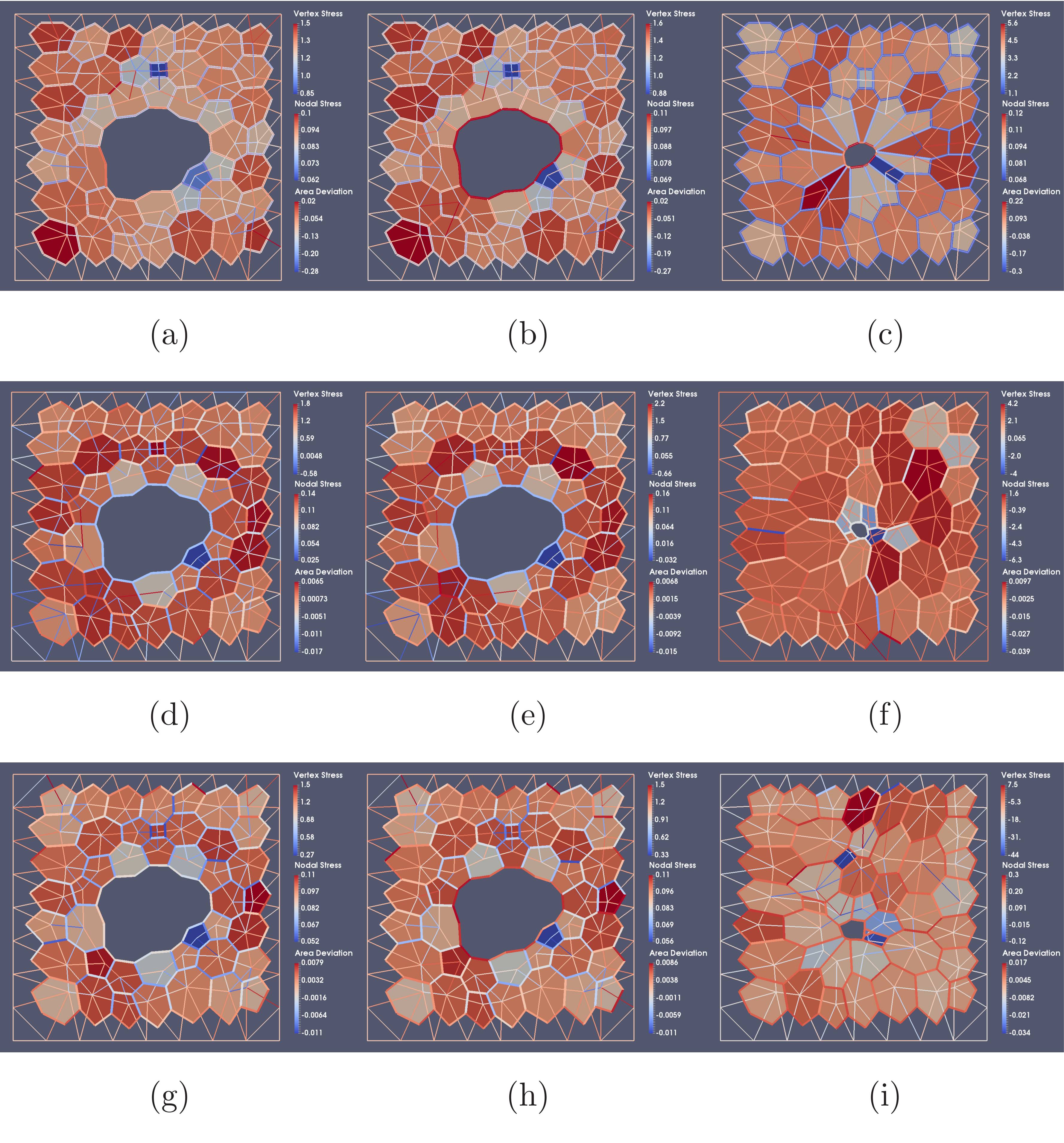}
\caption{Wound healing model visualised at different stages. The deviation from cells initial area, as well as traction values along nodal and vertex elements, are shown in the corresponding colour-bars  at each stage. (a-c) Wound healing in non-remodelling tissue. (d-f) Wound healing with full-network EPM. (g-i) Wound healing with split-network EPM. Figures (a), (d) and (g) correspond to time-steps just after wounding. Figures (b), (e) and (h) correspond to when extra contractility was applied on the elements at the wound ring. Figures (c), (f) and (i) correspond to when extra contractility at the wound edge caused wound closure. Corresponding movies of the simulations may be found in the Supplementary Material}
\label{wound}
\end{figure}

Wounding and wound healing processes are simulated during the consecutive steps below:  
\begin{enumerate}
\item To resemble the initial condition of \emph{in-vivo} tissue before wounding, the modelled tissue is let to reach a contractile state given by the values of $\veps^c_D$ and $\veps^c_V$ in Table \ref{t:par} and the evolution law affecting elements resting lengths, during 50 time-steps. This time is found to be sufficient to reach a steady asymptotic state.
\item Wounding by laser ablation of cells is analogised by a significant reduction of stiffness in nodal and vertex elements encircled by the wound edge,  as well as removing the area constraint on wounded cells. In wounded areas we set $k_D^{wounded}=0.1 k_D$  and $k_V^{wounded}=0.1 k_V$. Also, vertices at the wound edge are allowed to relax by resorting to the $\bxi$-relaxation. This is  done to avoid unrealistic zig-zag effects on the profile of the wound edge. Figures \ref{wound}a \ref{wound}d and \ref{wound}g show the tissue initially after wounding, without remodelling, and with full- and split-network remodelling, respectively. 
\item To simulate tissue eventual response to wounding, after 12 time-steps, contractility on the elements of the vertex network surrounding the wound (wound ring) is multiplied by 5 in order to pattern actomyosin concentration, as it has been experimentally tested \cite{brugues14}. Figures \ref{wound}b and \ref{wound}e show how the extra contractility on the wound edge results in higher tractions on the wound ring, at both non-remodelling and remodelling tissues.
\item Additional tractions on the wound ring cause the wounded area being squeezed by the cells on the wound boundary. Figures \ref{wound}c and \ref{wound}f show the wound closure with and without remodelling. Including remodelling during the tissue evolution results in less cell elongation at the wound edge and allows cells to relocate during wound closure. 
\end{enumerate}

In the full-network strategy (Figures \ref{wound}d-f), since the total residual of nodal and vertex networks were preserved at the nodes, the interplay of stresses in nodal and vertex networks could not preserve the higher stress in the vertex elements at the wound ring. Instead, the split-network strategy could provide the expected higher stress in the elements at the wound ring. This is due to preserving nodal residual independently in each of the networks. 

\clearpage

\section{Conclusions}\label{s:conclusions}

We have presented a hybrid cell-centred and vertex discretisation for biological tissues. This approach allows to independently control the material properties of the cell-boundaries and the cytoplasm (cell interior). The methodology  solves the mechanical equilibrium of the two networks in a coupled manner, and it has been shown that can reproduce relevant phenomena such as tissue extension or wound healing. 

The method resorts to a rheological law that is based on an evolution law of the resting length \cite{doubrovinsky17,munoz13b}. This evolution is controlled through the remodelling rate $\gamma$. For high values of $\gamma$, the tissue relaxes and adapts its reference free configuration rapidly, while for very low values of $\gamma$, an purely elastic response is recovered. 

The variations of the resting lengths allow also to design an Equilibrium-Preserving Map (EPM) that computes a set of resting lengths and traction field that mimics the force distribution on the nodal and vertex network before remodelling. The numerical examples presented show that this recovery of tractions alters minimally the stress state. 

We have just presented two-dimensional examples, but a three dimensional extension does not involve substantial changes neither in the hybrid approach and in the EPM strategy, if the vertex mechanics is maintained along bar elements. In case that mechanics at the cell boundaries is carried by the vertex faces, some additional modifications should be applied to the tractions and functional in the EPM.

The strategy described here opens also the possibility to energy decaying or methods where the actual reaction is relaxed in a controlled manner. This could be achieved by progressively reducing the nodal reaction used in the functional of the EPM. Also, the hybrid approach could be modified for handling cell proliferation or apoptosis (addition or removal of nodes). Current research is now being undertaken in this direction.

\section{Acknowledgements}

The authors gratefully acknowledge the financial support of the Spanish Ministry of Economy, Science and Competitiveness (MINECO) under grants DPI2013-32727-R and DPI2016-74929-R, and the Generalitat de Catalunya under grant 2014-SGR-1471. PM is also supported by the European Molecular and Biology Organisation (EMBO) under grant ASTF 351-2016.

\appendix
\section{Notation}\label{s:notation}

The notation used in this article is summarised in Tables \ref{t:notation} and \ref{t:notationb}.

\begin{table}[!htb]
\begin{center}
\begin{tabular}{|c|l|}
\hline
$A^m, A^m_0$ & Current and initial area of cell $m$. Eqn. \eqref{e:W_A}.\\
$A_T$ & Total area of cells on the tissue. Eqn. \eqref{e:A_T}.\\
$B^i$ & Vertices that surround node $i$. Eqn. \eqref{e:g_V}. \\
$\e^{ij}$ & Unit vector from node $\x^j$ to node $\x^i$. Eqn. \eqref{e:eij}\\
$\e^{IJ}$ & Unit vector from vertex $\y^J$ to vertex $\y^I$. Eqn. \eqref{e:eij}\\
$E_i, E_I$ & Error measures of vertex and nodal equilibrium, resp. Eqn. \eqref{e:Ei}\\
$\g_D^i$ & Nodal force contribution at node $i$. Eqn. \eqref{e:g_D} \\
$\g_V^i$ & Vertex force contribution at node $I$. Eqn. \eqref{e:g_V} \\
$\g_x$ & Residual vector stemming from $\frac{\partial W}{\partial \x}$. Eqn. \eqref{e:g}.\\
$\g_y$ & Residual vector stemming from $\frac{\partial W}{\partial \bxi}$. Eqn. \eqref{e:g}.\\
$\J$ & Matrix such that $(\y^I\times\y^J)\cdot\e_z=\y^I\cdot\J\y^J$. Eqn. \eqref{e:Ai}\\
$k_D, k_V$ & Stiffness of nodal and vertex network, resp. Eqn. \eqref{e:W_D} and \eqref{e:W_V}.\\
$l^{ij}, L^{ij}$ & Current and resting lengths of bar element between nodes \\
& $i$ and $j$. Eqn. \eqref{e:W_D}.\\
$l^{ij}, L^{IJ}$ & Current and resting lengths of bar element between vertices\\
&  $I$ and $J$. Eqn. \eqref{e:W_V}.\\
$\n^{ij}$ & Outward normal at vertex bar between vertices $I$ and $J$. \\
& Eqn. \eqref{e:Am2}. \\
$N_D$ & Number of bars in nodal network. Eqn. \eqref{e:W_D}.\\
$N_m$ & Number of segments that surround cell centered at $\x^m$. \\
& Eqn. \eqref{e:Am2}. \\
$N_{nodes}$ & Total number of nodes. Section \ref{s:ngeo}. \\
$\bar N_{nodes}$ & Number of internal nodes. Section \ref{s:vertex} \\
$N_{tri}$ & Total number of triangles in nodal network. Section \ref{s:ngeo}.\\
$N_V$ & Total number of vertex bars. Section \ref{s:vertex}.\\
$p^i(\bxi^I)$ & Shape function defining vertex positions. Eqn. \eqref{e:interp}.\\
$P^m$ & Set of segments that form boundary of cell $m$. Eqn \eqref{e:Am2}. \\
$q(\alpha)$ & Interpolation function of cell boundary. Eqn. \eqref{e:yq}.\\
\hline
\end{tabular}
\caption{Notation. The explicit definition of the symbols can be found in the indicated section or equation.}
\label{t:notation}
\end{center}
\end{table} 

\begin{table}[!htb]
\begin{center}
\begin{tabular}{|c|l|}
\hline
$\br^i$ & Nodal reaction due to boundary condition on node $i$. Eqn. \eqref{e:pi}. \\
$\br^i_D, \br_V^i$ & Nodal and vertex contribution to functional in EPM. Eqn. \eqref{e:r} \\
$S^i$ & Set of nodes connected to node $i$. Eqn. \eqref{e:g_D}\\
$S^I$ & Set of vertices connected to vertex $I$. Eqn. \eqref{e:g_V}\\
$\bt_D^{ij}$ & Traction vector at node $i$ exerted by nodal element $ij$. Eqn. \eqref{e:td}\\
$\bt_V^{IJ}$ & Traction vector at vertex $I$ exerted by vertex element $IJ$. \\
& Eqn. \eqref{e:tv}\\
$\mathcal T^I$ & Triangle where vertex $I$ is located. Section \ref{s:ngeo}.\\
$\T_n$ & Connectivity of nodal network at time $t_n$. Section \ref{s:ngeo}.\\
$W_A$ & Energy term associated to area penalisation. Eqn. \eqref{e:W_A}.\\
$W_D, W_D^{ij}$ & Total nodal strain energy and strain energy of nodal element $ij$.\\
& Eqn. \eqref{e:W_D}.\\
$W_V, W_V^{IJ}$ & Total vertex strain energy and strain energy of \\
& vertex element $IJ$. Eqn. \eqref{e:W_V}.\\
$W_\xi$ & Penalty term used in $\xi$-relaxation. Eqn. \eqref{e:Wxi} \\
$\x^i$ & Position of node $i$. Section \ref{s:ngeo}.\\
$\y^I$ & Position of vertex $I$. Section \ref{s:vertex}. \\
$\X_n$ & List of nodal positions at time $t_n$. Section \ref{s:ngeo}.\\
$\alpha$ & Local coordinate of points in vertex bars. Eqn. \eqref{e:yq}.\\
$\delta_{ij}, \delta_{IJ}$ & Kronecker delta. Eqn. \eqref{e:dAI}\\
$\delta_{mj}^{pq}, \delta_{IJ}^{PQ}$ & See definitions in eqn. \eqref{e:ghat}\\
$\Delta t$ & Time increment. Eqn. \eqref{e:devol}. \\
$\veps^c$ & Contractility of bar elements. Eqn. \eqref{e:evol}. \\
$\veps^c_D, \veps^c_V$ & Contractility employed in nodal and vertex network. Section \ref{s:wound}. \\
$\veps^{ij}, \veps^{IJ}$ & Strain at nodal and vertex bar elements, resp. Eqn. \eqref{e:W_D} and \eqref{e:W_V}. \\
$\gamma$ & Remodelling rate in rheological model. Eqn \eqref{e:ldot}.\\
$\lambda_A$ & Penalty terms for area constraint. See eqn. \eqref{e:W_A}.\\
$\lambda_\xi$ & Penalty terms for $\xi$-relaxation. See eqn. \eqref{e:Wxi}.\\
$\pi_F, \pi_S$ & Functionals of full- and split-network EPM. Eqn. \eqref{e:pi} and \eqref{e:piS}\\
$\theta^{ij}, \theta^{IJ}$ & Inverse of resting lengths $L^{ij}$ and $L^{IJ}$, resp. \eqref{s:full}.\\
$\bxi^I$ & Local coordinate of vertex $I$ in triangle $\mathcal T^I$.\\
\hline
\end{tabular}
\caption{Notation (continuation).}
\label{t:notationb}
\end{center}
\end{table} 

\section{Linearisation}\label{s:linearisation}

\subsection{General linearisation steps with $\xi$-relaxation}\label{s:lin_xi}

When $\xi$-relaxation is included, the total residual vector $\g=\{\g_x^T\  \g_y^T\}^T$ is split in a nodal $\g_x$ and $\xi$ contributions $\g_y$ (see equation \eqref{e:g}). Each nodal and vertex contribution is given by
\begin{align*}
\g^i_x&=\g^i_D+\g^i_V+\g^i_A,\\
\g^I_y&=\g^I_V+\g^i_A+\g^I_\xi.
\end{align*}

Vectors $\g^i_D$, $\g^i_V$ and $\g^i_A$ are written in equations \eqref{e:g_D}, \eqref{e:g_V} and \eqref{e:g_A}, and the vertex contributions  $\g^I_V$, $\g^I_A$ and $\g^I_\xi$ given in equations \eqref{e:g_xi}. The non-linear equations $\g=\bO$ are solved with a Newton-Raphson process that at each iteration $k$ reads
\begin{align}\label{e:NR}
\left\{\begin{array}{c}
\delta\x\\
\delta\bxi
\end{array}\right\}
=-\left[\begin{array}{cc}
\K_{xx} & \K_{xy} \\
\K_{y x} & \K_{yy} 
\end{array}\right]_k^{-1}
\left\{\begin{array}{c}
\g_x\\
\g_y
\end{array}\right\}_k
\end{align}
and is updated as
\begin{align*}
\left\{\begin{array}{c}
\x\\
\bxi
\end{array}\right\}_{k+1}
=\left\{\begin{array}{c}
\x\\
\bxi
\end{array}\right\}_k
+\left\{\begin{array}{c}
\delta\x\\
\delta\bxi
\end{array}\right\}
\end{align*}
 as long as the two following conditions are met,
\begin{align*}
\left\{\begin{array}{rl}
\sqrt{\left \|\delta\textbf{\textit{x}}\right\|^2+\left\|\delta\bxi\right\|^2}&>tol \\ 
\left\|\textbf{\textit{g}}\right\|&>tol
\end{array}\right.
\end{align*}
 with $tol$ a sufficiently small tolerance. In our numerical examples we used $tol=1e-10$.

The block matrices in \eqref{e:NR} correspond to the following linearisation terms,
\begin{align}\label{e:KNR1}
\K_{xx}^{ij} &=\frac{\partial \g^i_D}{\partial \x^j} +\frac{\partial \g^i_V}{\partial \x^j} + \frac{\partial \g^i_A}{\partial \x^j}\\
\K_{xy}^{iJ}&=\phantom{\frac{\partial \g^i_D}{\partial \bxi^J}+} \frac{\partial \g^i_V}{\partial \bxi^J} + \frac{\partial \g^i_A}{\partial \bxi^J}\label{e:KNR2} \\
\K_{y x}^{Ij}&=\frac{\partial \g^I_V}{\partial \x^j} +\frac{\partial \g^I_A}{\partial \x^j} \nonumber\\
\K_{yy}^{IJ}&=\frac{\partial \g^I_V}{\partial \bxi^J} +\frac{\partial \g^I_A}{\partial \bxi^J} + \frac{\partial \g^I_\xi}{\partial \bxi^J} \label{e:KNR3}
\end{align}
where due to the expressions of $\g_D^i$ and $\g_\xi^I$, we have used the fact that $\frac{\partial \g^i_D}{\partial \y^J}$ and $\frac{\partial \g^I_\xi}{\partial \x^j}$ vanish. Also note that since our equlibrium equations stem from the linearisation of an energy function $W(\x, \bxi)$, we have that
\[
\K_{xy}^{iJ}=\frac{\partial^2 (W_V+W_A)}{\partial \x^i\partial\bxi^J}=\left[\frac{\partial^2 (W_V+W_A)}{\partial \bxi^I\partial \x^j}\right]^T={\K_{yx}^{Ij}}^{T}.
\]
In the next sections we will give the linearisation of the terms in \eqref{e:KNR1}-\eqref{e:KNR3}.

\subsection{Linearisation of nodal and vertex tractions $t_D$ and $t_V$}

Many of the derivations detailed below will involve the linearisation of the traction vectors given in \eqref{e:tdtv},
\begin{align*}
\bt_D^{ij}&=\frac{\partial W_D^{ij}}{\partial\x^i}
=\frac{\veps^{ij}}{L^{ij}}\left(1-\frac{l^{ij}}{L^{ij}}\frac{\partial L^{ij}}{\partial l^{ij}}\right)\e^{ij}\\
\bt_V^{IJ}&=\frac{\partial W_V^{IJ}}{\partial\y^I}
=\frac{\veps^{IJ}}{L^{IJ}}\left(1-\frac{l^{IJ}}{L^{IJ}}\frac{\partial L^{IJ}}{\partial l^{IJ}}\right)\e^{IJ}
\end{align*}

The factor $\frac{\partial L}{\partial l}$ is zero when the resting length is constant, but for the rheological law presented in Section \ref{s:rheo}, this factor is given in equation \eqref{e:dLdl}. In the subsequent expressions we will need the derivatives of the traction vectors above. We define matrix
\begin{subequations}\label{e:Kt}
\begin{align}
\K^{ii}_t:=\frac{\partial \bt_D^{ij}}{\partial \x^i}=-\frac{\partial \bt_D^{ji}}{\partial \x^i}=-\K^{ji}_t=-\K^{ij}_t=\K^{jj}_t
\end{align}
which after making use of \eqref{e:deps}, it can be deduced that
\begin{align}
\begin{aligned}
\K^{ij}_t&=(-1)^{\delta_{ij}+1}\left(\left(a^{ij}a^{ij}-\frac{\veps^{ij}}{l^{ij}} a^{ij}+\veps^{ij}b^{ij}\right)\e^{ij}\otimes\e^{ij}
+\frac{\veps^{ij}a^{ij}}{l^{ij}}\mathbf I\right)\\
a^{ij}&=\frac{1}{L^{ij}}\left(1-\frac{l^{ij}}{L^{ij}}\frac{\partial L}{\partial l}\right)\\
b^{ij}&=\frac{1}{L^{ij}}\frac{\partial L}{\partial l} \left(-a^{ij}+\frac{1}{L^{ij}}\left(\frac{l^{ij}}{{L^{ij}}^2}-1\right)\right)
\end{aligned}
\end{align}
\end{subequations}

A similar derivation is obtained for $\frac{\partial \bt_V^{IJ}}{\partial \y^I}$, but replacing $ij$ by $IJ$. In this case, we also note that from the interpolation in \eqref{e:interp} we have,
\begin{align*}
\frac{\partial \bt_V^{IJ}}{\partial \x^j}&=\K^{IJ}_t\left(\frac{\partial \y^J}{\partial \x^j}-\frac{\partial\y^I}{\partial\x^j}\right)
=\K_t^{IJ}\left(p^j(\bxi^J)-p^j(\bxi^I)\right) \\
\frac{\partial \bt_V^{IJ}}{\partial \bxi^J}
&=\frac{\partial \bt_V^{IJ}}{\partial \y^I}\frac{\partial \y^I}{\partial \bxi^J}
+\frac{\partial \bt_V^{IJ}}{\partial \y^J}\frac{\partial \y^J}{\partial \bxi^J}
=\K^{IJ}_t\sum_{\x^j\in\mathcal T^J}\x^j\otimes\nabla p^j(\bxi^J)
\end{align*}
where $p^i(\bxi^I)=0$ if $i\notin\mathcal T^I$.

\subsection{Linearisation terms in $\K^{ij}_{xx}$}

By using the expressions of $\g_D^i$, $\g_V^i$ and $\g_A^i$ in \eqref{e:g_D}, \eqref{e:g_V} and \eqref{e:g_A}, and the definition of $\K_t^{ij}$ in \eqref{e:Kt}, it can be deduced that
\begin{align*}
\frac{\partial \g^i_D}{\partial \x^j}&=\sum_{j\in S^i}\K_t^{ij}\\
\frac{\partial \g^i_V}{\partial \x^j}&=\sum_{I\in B^i}\sum_{J\in S^I}\K_t^{IJ}\left(p^j(\bxi^J)-p^j(\bxi^I)\right)\\
\frac{\partial \g^i_A}{\partial \x^j}&=\frac{\lambda_A}{2}\J\!\!\sum_{m\in \bar S^i}\left(A^m-A^m_0\right)\sum_{IJ\in P ^m}\left(p^i(\bxi^I)p^j(\bxi^J)-p^i(\bxi^J)p^j(\bxi^I)\right)\\
+&\frac{\lambda_A}{4}\!\!\sum_{m\in \bar S^i}
\sum_{IJ\in P^m}\!\!\J\left(p^i(\bxi^I)\y^J-p^i(\bxi^J)\y^I\right)\!\!\otimes\!\!\!\!\!\!
\sum_{KL\in P^m}\!\!\J\left(p^j(\bxi^K)\y^L-p^j(\bxi^L)\y^K\right)
\end{align*}

\subsection{Linearisation terms in $\K^{iJ}_{xy}$}

From the expressions of $\g_V^i$ and $\g_A^i$ in \eqref{e:g_V} and \eqref{e:g_A}, and from equation \eqref{e:dydx},  it can be also deduced that

\begin{align*}
\frac{\partial \g^i_V}{\partial \bxi^J}&
=\left(\sum_{K\in S^J}\bt^{JK}_V\right)\otimes\nabla p^i(\bxi^J)
+\sum_{I\in B^i}p^i(\bxi^I)\sum_{J\in S^I}\K^{IJ}_t\frac{\partial\y^J}{\partial\bxi^J}\\
\frac{\partial \g^i_A}{\partial \bxi^J}&=
\frac{\lambda_A}{2}\J\!\!\sum_{m\in \bar S^i}\left(A^m-A^m_0\right)\sum_{IJ\in P ^m}^{N_m}\left(
p^i(\bxi^I)\frac{\partial\y^J}{\partial\bxi^J}-\y^I\otimes \nabla p^i(\bxi^J)
\right)\\
&+\frac{\lambda_A}{2}\J\!\!\sum_{m\in \bar S^i}\sum_{IJ\in P ^m}^{N_m}\left(
p^i(\bxi^I)\y^J-p^i(\bxi^J)\y^I\right)\otimes
\frac{\partial A^m}{\partial\bxi^J}
\end{align*}
with $\frac{\partial A^m}{\partial\bxi^J}$ give in \eqref{e:dAI}.

\subsection{Linearisation terms in $\K^{IJ}_{yy}$}

The linearisation of $\g_V^I$, $\g_A^I$ and $\g_\xi^I$ in \eqref{e:g_xi} yields
\begin{align*}
\frac{\partial \g^I_V}{\partial \bxi^J}&= \sum_{K\in S^I}\sum_{i\in\mathcal T^I}
\left(\nabla p^i(\bxi^I)\otimes\x^i\right)\left(\K_t^{II}\delta_{IJ}\frac{\partial \y^I}{\partial\bxi^J}+\K_t^{IK}\delta_{KJ}\frac{\partial\y^K}{\partial\bxi^J}\right)\\
\frac{\partial \g^I_A}{\partial \bxi^J}&= 
\lambda_A\sum_{m=1}^{\bar N_{nodes}} \frac{\partial A^m}{\partial\bxi^I}\otimes\frac{\partial A^m}{\partial\bxi^J}
+\lambda_A \sum_{m=1}^{\bar N_{nodes}}
\left(A^m-A^m_0\right)\frac{\partial^2 A^m}{\partial\bxi^I\partial\bxi^J}\\
\frac{\partial \g^I_\xi}{\partial \bxi^J}&=\lambda_\xi\delta_{IJ}\mathbf I
\end{align*}
where the expressions of $\frac{\partial\y^I}{\partial\bxi^I}$ and $\frac{\partial A^m}{\partial\bxi^I}$ are given in \eqref{e:dydx} and in \eqref{e:dAI}, respectively, and 
\[
\frac{\partial^2 A^m}{\partial\bxi^I\partial\bxi^J}=\sum_{KL\in P^m} \left(
\delta_{KI}\delta_{LJ}\left(\frac{\partial\y^K}{\partial\bxi^I }\right)^T
\J\frac{\partial\y^L}{\partial \bxi^J}
-\delta_{LI}\delta_{KJ}\left(\frac{\partial\y^L}{\partial\bxi^I }\right)^T
\J\frac{\partial\y^K}{\partial \bxi^J}\right).
\]


\bibliographystyle{wileyj}

\end{document}